\documentclass[aps,prl,twocolumn,superscriptaddress,floatfix]{revtex4-2}
\usepackage{amssymb}
\usepackage{physics}
\usepackage{siunitx}
\usepackage{graphicx}
\usepackage{amsmath}
\usepackage{amsthm}
\usepackage{amsfonts}
\usepackage[T1]{fontenc}
\usepackage{array}
\usepackage{multirow}
\usepackage{color}
\usepackage{esint}
\usepackage{bm}
\usepackage{color}
\usepackage{bbm}
\usepackage{hyperref}
\usepackage{babel}

\newcommand{\edit}[1]{{\color{red} {#1}}}
\newcommand{\todo}[1]{{\color{cyan} {#1}}}

\newcommand{\lettersection}[1]{\emph{\todo{#1}}.---}
\newcommand{\moire}{moir\'e\ }
\usepackage{hyperref}
\hypersetup
{	colorlinks,%
	citecolor=green,%
	linkcolor=red,%
	urlcolor=blue%
}
\allowdisplaybreaks[4]
\begin{document}

\global\long\def\id{\mathbbm{1}}
\global\long\def\ui{\mathbbm{i}}
\global\long\def\ud{\mathrm{d}}

\title{Impurity-induced thermal crossover in fractional Chern insulators}

\author{Ke Huang}
\affiliation{Department of Physics, City University of Hong Kong, Kowloon, Hong Kong SAR, China}

\author{Sankar Das Sarma}
\affiliation{Condensed Matter Theory Center and Joint Quantum Institute, University of Maryland, College Park, Maryland 20742, USA}
\affiliation{Kavli Institute for Theoretical Physics, University of California, Santa Barbara, California 93106, USA}

\author{Xiao Li}
\email{xiao.li@cityu.edu.hk}
\affiliation{Department of Physics, City University of Hong Kong, Kowloon, Hong Kong SAR, China}

\date{\today}

\begin{abstract}
The recent experimental observation of fractional quantum anomalous Hall (FQAH) states in rhombohedral multilayer graphene has attracted significant attention. 
One of the most intriguing observations is that the FQAH states at various fractional fillings give way to IQAH states as the temperature is lowered. 
In this work, we propose a mechanism for the appearance of FQAH states within a finite temperature range in a toy model. 
The model consists of a flat Chern band and impurities, and we analyze the effects of impurities on the system's behavior at finite temperatures. 
Because a hole-doped FQAH state has a large entropy due to the abundant quasiparticle excitations, we believe that the crossover may arise from the competition between the energy penalty for thermal excitations and the increase in entropy. 
We support our theoretical argument with numerical calculations using exact diagonalization. 
Our results suggest that impurities may play a crucial role in the crossover from the FQAH to IQAH states in rhombohedral pentalayer graphene. 
\end{abstract}

\maketitle

\lettersection{Introduction}
The search for exotic quantum matter with intrinsic topological order has been a central focus in condensed matter physics over the past few decades. 
One such example is the fractional Chern insulator (FCI), which serves as an analog of the fractional quantum Hall effect in the absence of any applied magnetic field. 
Some of the earliest proposals for the FCI were made in the context of a flat Chern band with strong interactions~\cite{Tang2011,Sun2011,Neupert2011,Sheng2011,Regnault2011}. 
However, the experimental realization of FCI states without an applied magnetic field, now commonly known as the fractional quantum anomalous Hall (FQAH) states, remained elusive until recently. 
Several research groups have now reported the observation of the FQAH states in various systems, including twisted bilayer MoTe$_2$~\cite{Cai2023,Zeng2023,Park2023,Xu2023} and rhombohedral multilayer graphene~\cite{Lu2024,XieJian2025,Waters2025,Lu2025,Choi2025}. 
These groundbreaking experiments have opened new avenues for the study of topological states of matter.

Meanwhile, these experimental breakthroughs have raised new questions regarding the observed FQAH states, especially in rhombohedral pentalayer graphene (PLG).  
A particularly puzzling finding, as reported recently by Lu et al.~\cite{Lu2025}, is that the FQAH states at some fractional fillings cross over to integer quantum anomalous Hall (IQAH) states as the temperature is lowered. 
Such a crossover also depends on the applied displacement field, which can control the band structure and screen the impurities in the sample. 
This finding is quite unexpected since the FQAH states are fragile and expected to be stable at extremely low temperatures~\cite{LuHongYu2024}. 
Therefore, it has attracted immediate attention in the community~\cite{DasSarma2024}, although a comprehensive understanding of this unexpected crossover is still lacking. 

In this Letter, we attempt to provide a quantitative theory to explain why the FQAH effect could only be stabilized within a finite temperature range,  $T_e < T <T_\text{FCI}$, where $T_\text{FCI}$ is the temperature scale associated with the FCI gap or other low-energy excitations~\cite{LuHongYu2024}. 
Above $T_\text{FCI}$, various excitations are expected to destroy the FQAH effect. 
The key new ingredient in our theory is the introduction of a lower temperature scale, $T_e>0$. 
We argue that, as the temperature is lowered, carriers can be localized by impurities, reducing the ``active'' carriers in the topological flat band that can form the FQAH state {and leading to a hole-Wigner crystal (WC) with nonvanishing Hall conductivity}. 
At $T<T_e$, the effective filling factor is reduced to a value insufficient to maintain the FQAH state, thus destroying the FQAH effect. 
As long as $T_e < T_\text{FCI}$, the mechanism proposed by our theory could occur in the experiment. 
Our arguments will be supported by explicit numerical calculations using a toy model with impurities. 
We emphasize that the achievement of $T_e<T_\text{FCI}$ in this work is due to a unique feature of the FCI state: the large entropy arising from the abundant quasiparticle excitations.

In the remainder of this Letter, we first present the toy model and our theoretical argument for the thermal crossover.
In particular, we will carefully define the temperature scales $T_e$ and $T_\text{FCI}$ and explain the mechanism for the crossover.  
We then support our theory with explicit numerical calculations. 
We finally conclude with a discussion of the implications of our theory for the experimental situations. 

\begin{figure}[t]
	\center
	\includegraphics[width=\columnwidth]{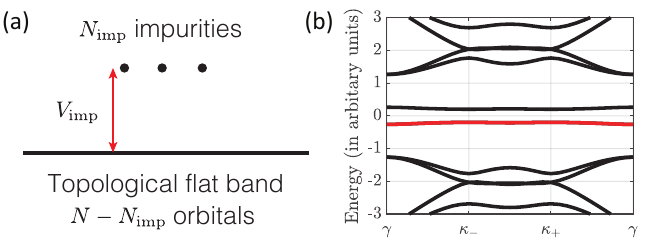}
	\caption{\label{Fig:diagram} 
	(a) Illustration of the toy model with impurities. The total system size is $N$, and $N_{\text{imp}}$ impurity orbitals are energetically separated from the topological flat band. 
	(b) Band structure for CTBG with $w=0.9w_{\text{chiral}}$. We project the Hamiltonian to the highest valence band (the red line).
	}
\end{figure}

\lettersection{Toy model with impurities}
The essential ingredients of our toy model include a flat Chern band and impurities, as sketched in Fig.~\ref{Fig:diagram}(a). 
We start by considering a toy model featuring a well-isolated and flat Chern band that is fractionally filled. 
We then assume that the whole system can be projected to this flat band and that the band dispersion can be ignored. This assumption is valid when all other energy scales, including interaction, impurities, and temperature, are much smaller than the gaps between the flat band and other bands, but substantially larger than the width of the flat band.
As a result, we only need to consider the subspace spanned by the $n_e$-particle states of the flat band $\prod_{i=1}^{n_e}f_{\vb*k_i}^\dag\ket{0}$, where $f_{\vb*k}^\dag$ is the creation operator of a particle with momentum $\vb*k$ in the flat band.
Consequently, the complete model Hamiltonian reads as
\begin{align}\label{Eq:Hamiltonian}
	H&=P_{\text{flat}}(H_{\text{int}}+H_{\text{imp}})P_{\text{flat}},
\end{align}
where $P_{\text{flat}}$ is the projection operator of the subspace, and $H_{\text{int}},H_{\text{imp}}$ are the interaction and impurity potential, respectively.

To put our discussion into a concrete example, we consider the highest valence band of chiral twisted-bilayer graphene (CTBG)~\cite{Tarnopolsky2019}. The continuum model Hamiltonian for CTBG is given by
\begin{align}\label{Eq:CTBG}
	&H_{\text{CTBG}}=\left[\begin{array}{cc}\gamma \vb*{\sigma}_{\theta/2}\vdot \hat{\vb*k}_- & T(\hat{\vb*r}) \\ T^\dag(\hat{\vb*r}) & \gamma  \vb*{\sigma}_{-\theta/2}\vdot \hat{\vb*k}_+ \end{array}\right]+\Delta\mathbb{I}\otimes\sigma_z,
\end{align}
The operators $f_{\vb*k}^\dag$ then lives in this flat band. 
In the above equation, $\gamma$ is the Fermi velocity of monolayer graphene, $\theta$ is the twist angle between the two graphene sheets, $\vb*{\sigma}_{\theta}=e^{-i\theta \sigma_z/2}\vb*\sigma e^{i\theta\sigma_z/2}$, and $\Delta$ is used to split the conduction and valence bands. 
In addition, $\hat{\vb*k}_\pm=\hat{\vb* k}-\vb*\kappa_\pm$ are the momentum operators of the top and bottom layers, and $\vb*\kappa_\pm=k_\theta(-\sqrt{3}/2,\pm 1/2)$ are the $K$ valleys of the top and bottom layer. Here, we take $\theta=1.1^\circ$ and define 
$k_\theta=8\pi\sin(\theta/2)/(3a)$, 
where $a$ is the lattice constant of graphene. The interlayer coupling in CTBG is given by
\begin{align}
	T(\hat{\vb*r})=T_0+T_1e^{-i \vb*g_1\vdot \hat{\vb*r}}+T_2 e^{-i \vb*g_2\vdot \hat{\vb*r}},
\end{align}
where $T_n=w[\sigma_x\cos(2\pi n/3)+\sigma_y\sin(2\pi n/3)]$, and the two reciprocal vectors of the \moire Brillouin zone (MBZ) are given by 
\begin{align*}
	\vb*g_1=k_\theta(\sqrt{3}/2,3/2), \; 
	\vb*g_2=k_\theta(-\sqrt{3}/2,3/2). 
\end{align*}

The lowest conduction band and the highest valence band of CTBG are separated by a gap of $2\Delta$. 
They become exactly flat and can be mapped to the lowest Landau level (LLL) dressed by real space modulation~\cite{Wang2021} when $w=w_{\text{chiral}}\approx0.586\gamma k_\theta$. 
We take $w=0.9w_{\text{chiral}}$ in the main text because the interaction renormalizes the band dispersion for the hole-doping case~\cite{Abouelkomsan2020}, and we find that tuning away from the exact flat-band limit helps improve the robustness of the FCI state. 
In the supplemental material (SM)~\cite{supplement} (including references~\cite{Platzman1993,Dong2023c}), we show that the precise value of $w$ does not play an essential role in our theory as long as $w\approx w_{\text{CTBG}}$. 
Moreover, the two bands remain very flat for $w=0.9w_{\text{chiral}}$, as shown in Fig.~\ref{Fig:diagram}(b). 
Therefore, the flat band limit is still legitimate.
Additionally, the one-band projection is justified despite the spin-valley degeneracy in CTBG, because, within the filling range $0 \leq \nu \leq 1$ studied in this work, Coulomb interactions induce spontaneous symmetry breaking that polarizes the system into a single spin-valley flavor~\cite{Nuckolls2020,Lian2021}. 
In what follows, we will set the energy of the flat band to be zero, because the absolute energy scale of Eq.~\eqref{Eq:CTBG} (and thus the value of $\gamma$) is irrelevant to Eq.~\eqref{Eq:Hamiltonian} in the flat-band limit.

We further consider an extensively used many-body interaction in twisted-bilayer graphene systems~\cite{Bernevig2021,Christos2022,Kwan2023}, 
\begin{align}
	H_{\text{int}}=\frac1{2A}\sum_{\vb*q}V(q)(\rho_{\vb*q}-\delta_{\vb*q,0}/2)(\rho_{-\vb*q}-\delta_{\vb*q,0}/2),
\end{align}
where $\rho_{\vb*q}=\sum_{\vb*k}c^\dag_{\vb*k+\vb*q}c_{\vb*k}$, and $c_{\vb*k}$ is the annihilation operator of the plane wave, $A$ is the area of the system, and we take the bare Coulomb interaction $V(q)=\lambda/(k_\theta q)$ with $\lambda=1$ set as the energy unit. 
Finally, we consider an impurity potential with $N_{\text{imp}}$ delta functions, 
\begin{align}
	H_{\text{imp}}=V_{\text{imp}}\sum_{i=1}^{N_{\text{imp}}}\delta(\hat{\vb*r}-\vb*r_i)/\expval{n(\vb*r_i)}_{\text{full}}.
\end{align}
As the valence band has a nonuniform real-space density distribution, we normalize the delta potential by the local density of the completely filled valence band $\expval{n(\vb*r_i)}_{\text{full}}$. 
Consequently, the impurity strength $V_{\text{imp}}$ has the dimension of energy. 
The explicit forms of the projected Hamiltonian are given in the SM~\cite{supplement}.

\lettersection{Mechanism for the crossover}
We now explain the mechanism for the crossover from pinned hole-WCs to FCI states at finite temperatures. 
Specifically, we consider the filling factor $\nu$ to be slightly greater than $2/3$, or equivalently, the hole-filling factor $\nu_h:=1-\nu$ to be slightly less than $1/3$. 
At zero temperature and in the absence of impurities, this model is known to be an FCI at $\nu_h=1/3$~\cite{Wang2021}.  
However, impurities with an energy $E_\text{imp}>0$\edit{$V_\text{imp}>0$} trap holes, reducing the number of ``active'' holes in the flat band that can form the FCI state. 
For a strong impurity potential, the number of active particles in the flat band at zero temperature is $n_h - N_\text{imp}$, where $n_h=N-n_e$ denotes the total number of holes, and $N$ is the system size. 
While an FCI state is still stable when the hole-filling factor slightly deviates from $1/3$, if there are too many impurities, $n_h - N_\text{imp}$ may fall below the threshold needed to maintain the FCI state.
Particularly, the long-range interaction tends to produce a hole-WC pinned by impurities at low hole fillings, and as the hole-WC does not contribute to the conductivity, the Hall conductivity is identical to that of a completely filled topological band, leading to the IQAH effect.
The situation changes at finite temperatures, however. 
Thermal excitations from the impurity orbitals to the flat band are now possible if they lower the system's total free energy. 
These excited holes can increase the number of active holes in the flat band above the threshold, potentially restoring the FCI state. 

The above heuristic argument can be made more concrete by estimating the free energy of the system, which allows us to determine the temperature scale $T_e$ at which the crossover occurs.
Let us consider a scenario where, at a finite temperature $0<T<T_\text{FCI}$, $x$ holes (where $x < N_\text{imp}$) are excited from the impurity orbitals to the flat band. 
Because $n_h-N_{\text{imp}}+x < N/3$, the system contains $N_{\text{qh}}(x)$ quasiparticle excitations. 
These excitations constitute a low-energy manifold that is separated from the high-energy states by an energy gap $\Delta_\text{FCI}$~\cite{Regnault2011}, which is closely related to $T_{\text{FCI}}$, although other low-energy excitations may also affect $T_\text{FCI}$~\cite{LuHongYu2024}. 
The total entropy of the system is approximately given by $\ln\qty[\binom{N_\text{imp}}{x}N_{\text{qh}}(x)]$, where $\binom{N_\text{imp}}{x}$ represents the number of ways to excite the particles from the impurities to the flat band and $N_{\text{qh}}(x)$ represents the ways of forming an FCI state after the excitation. 
Hence, the free energy of the system is approximated by
$F\approx xV_{\text{imp}} - T\ln\qty[\binom{N_\text{imp}}{x}N_{\text{qh}}(x)]$, 
where the first term is the energy penalty for the thermal excitation.  
The condition that $F \leq 0$ defines an excitation temperature scale 
\begin{align}
	T_{e} \sim \frac{x V_{\text{imp}}}{\ln\qty[\binom{N_\text{imp}}{x}N_{\text{qh}}(x)]}. \label{Eq:T_e}
\end{align}
Therefore, a crossover to an FCI state at finite temperatures is possible if $T_e<T_{\text{FCI}}$. 

The above free energy argument and the criterion $T_e<T_{\text{FCI}}$ for the crossover are anticipated to be valid even for generic impurities. 
However, it is hard to estimate $T_e$ in the latter case because it is generally impossible to determine the energy penalty and the entropy. 
The advantage of our toy model is that the $T_{e}$ can be decreased by either increasing $\binom{N_\text{imp}}{x}$ or $N_{\text{qh}}(x)$, so the possibility of the crossover can be theoretically and numerically demonstrated.

\begin{figure}[t]
	\center
	\includegraphics[width=\columnwidth]{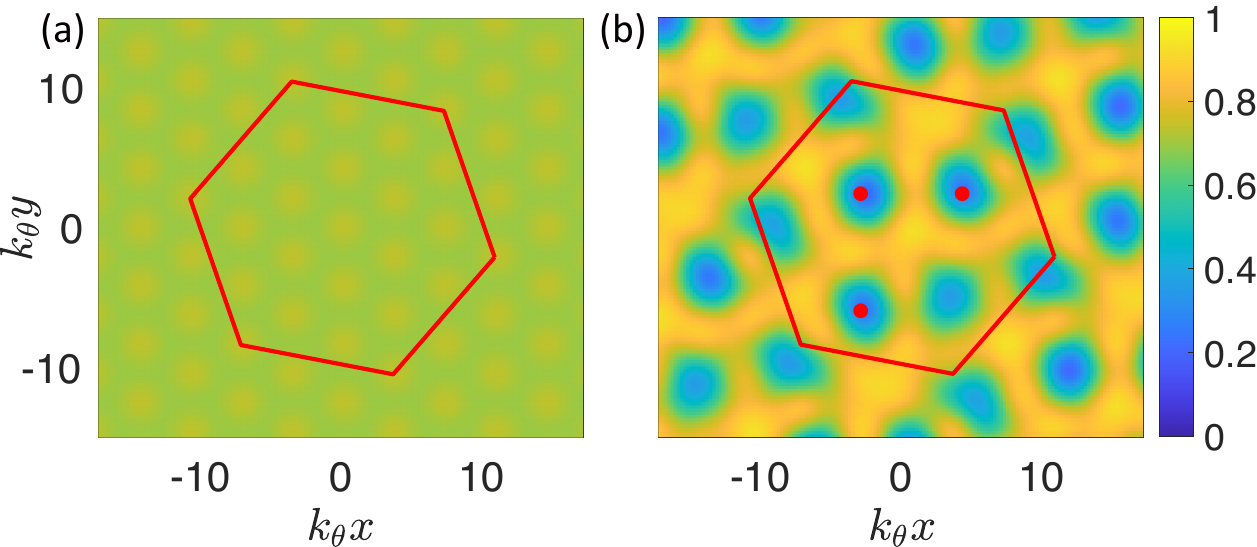}
	\caption{\label{Fig:density} 
	Relative density of the ground state for (a) $V_{\text{imp}}=0$ and (b) $V_{\text{imp}}=0.0035$. The color represents the relative density, the three red dots indicate the position of the three impurities, and the red hexagon delineates the periodic boundary of the system with $N=21$ unit cells and $n_h=6$ holes.
	}
\end{figure}

\lettersection{Numerical results}
We now numerically test our theory by calculating the finite-temperature density matrix of the toy model using numerical exact diagonalizations. 
To better observe the temperature-induced crossover from the hole-WC to the FCI phase, we need to suppress $T_e$ as much as possible by increasing either $\binom{N_\text{imp}}{x}$ or $N_{\text{qh}}$. 
However, it is difficult to increase $\binom{N_\text{imp}}{x}$ numerically because it necessitates an extremely large system size beyond the current capability of exact diagonalizations. 
Nonetheless, one can readily increase $N_{\text{qh}}$ by choosing a filling factor slightly less than the exact fractional filling.

We first investigate the impurity effects at zero temperature by studying the real-space density of the ground state. 
One hallmark of the fractional quantum Hall state is that it has a uniform real-space density, and the FCI state has a similar feature after removing the intrinsic periodic modulation of the band and considering the relative density 
$\expval{n(\vb*r)}_{\text{rel}}:=\expval{n(\vb*r)}/\expval{n(\vb*r)}_{\text{full}}$. 
By contrast, the holes in a hole-WC are localized and confined within their individual neighborhoods. 
In Fig.~\ref{Fig:density}, we calculate the relative density of the ground state in a system of $N=21$ unit cells (highlighted as a red hexagon) and $n_h=6$ holes with periodic boundary conditions, where the periodicity is defined by $\vb*a_1=\frac{4\pi}{3k_\theta}(2\sqrt{3},3)$ and $\vb*a_2=\frac{2\pi}{3k_\theta}(-5\sqrt{3},3)$. 
Without impurities, the ground state has a rather uniform relative density of $\expval{n(\vb*r)}_{\text{rel}}\approx 2/3$ [see Fig.~\ref{Fig:density}(a)], suggesting an FCI ground state. 
Upon introducing three impurities with $V_{\text{imp}}=0.0035$ [see Fig.~\ref{Fig:density}(b)], the six holes are concentrated within six pockets, three of which are located around the three impurities, indicating that the holes form a pinned hole-WC. 

\begin{figure}[t]
	\center
	\includegraphics[width=0.45\textwidth]{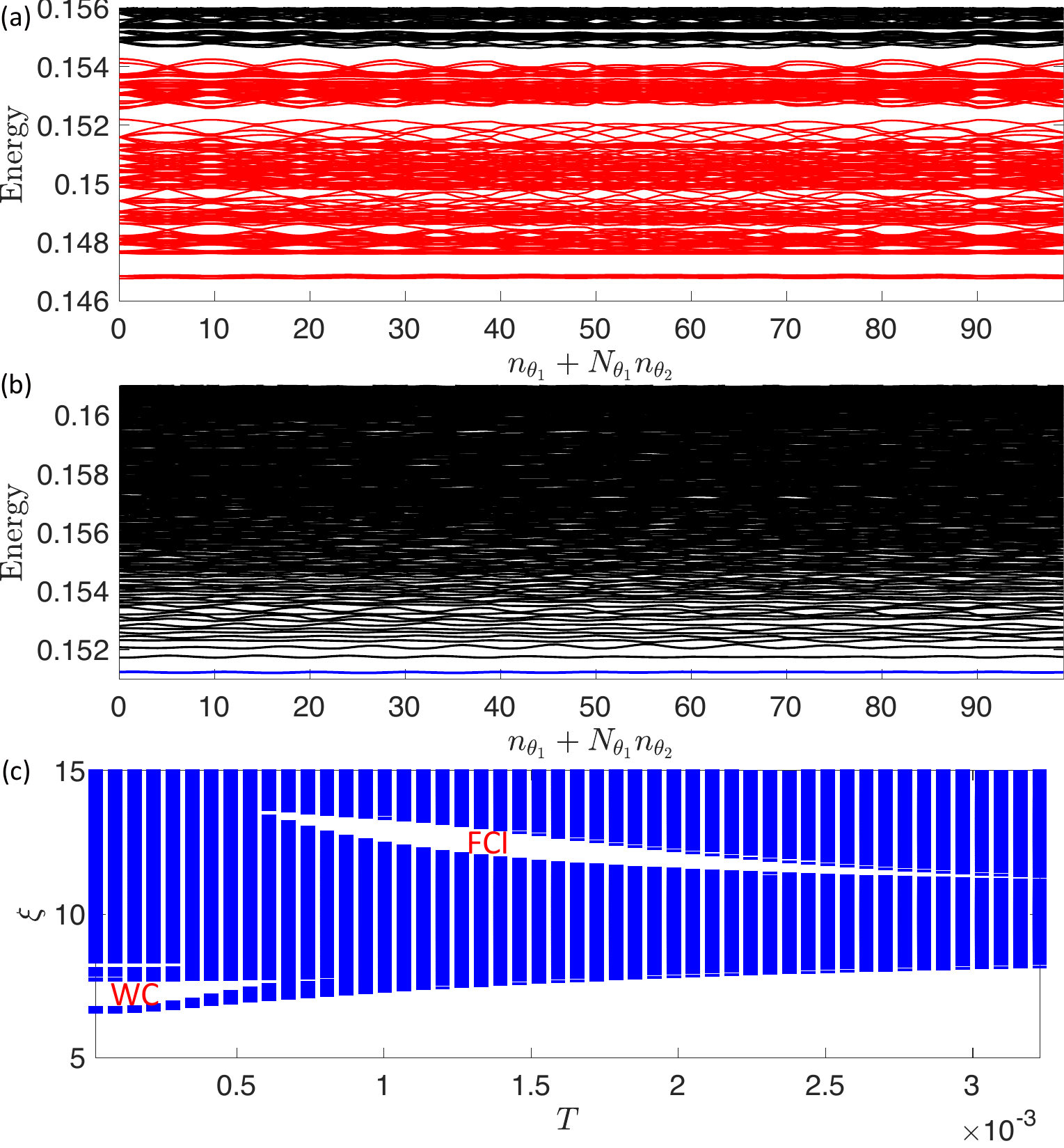}
	\caption{\label{Fig:Spectrum} 
	(a) and (b) are the energy spectrum as a function of the flux $n_{\theta_x} + N_{\theta_x}n_{\theta_y}$ at $V_{\text{imp}}=0$ and $V_{\text{imp}}=0.0035$, respectively.
	The red lines in (a) highlight the low-energy manifold of the FCI state (the $196$ quasiparticle excitations), and the blue line in (b) highlights the ground state of the pinned hole-WC.  
	Here we use $N_{\theta_1} = N_{\theta_2} = 10$ and take $\theta_j=n_{\theta_j}/N_{\theta_j}$ for $n_{\theta_j}=0,\cdots,N_{\theta_j}-1$. 
	(c) HES as a function of temperature at $V_{\text{imp}}=0.0035$. 
	For the entanglement gap at $T\lesssim 5\times 10^{-4}$, there are $20$ states below the gap, in agreement with the quasiparticle excitation of FL. 
	For the entanglement gap at $T\gtrsim 5\times 10^{-4}$, there are $637$ states below the gap, in agreement with the (1,3)-permissible quasiparticle excitation of FCI. 
	Here, we use the same system and impurities as the ones in Fig.~\ref{Fig:density}, and we take $n_a=3$ for the HES calculation.
	}
\end{figure}

We further verify this transition at zero temperature by analyzing the energy spectrum and the many-body Chern number. 
The twisted boundary condition is introduced by requiring the creation operators to satisfy $\mathcal{T}(\vb* a_j)f_{\vb* k}^\dag\mathcal{T}(\vb* a_j)^{-1}=e^{i\theta_j}f_{\vb* k}^\dag$, where $\mathcal{T}(\vb* x)$ is the translation operator.  
We then uniformly sample $N_{\theta_1} \times N_{\theta_2}$ values of $(\theta_1,\theta_2)$ in the range $[0,2\pi)\times[0,2\pi)$ and calculate the energy spectrum as a function of the twisted boundary condition in Fig.~\ref{Fig:Spectrum}.
In the clean limit, we verify that the system does contain a low-energy manifold of 196 states separated by an energy gap from the other states [Fig.~\ref{Fig:Spectrum}(a)], in agreement with the quasiparticle excitations of an FCI state. 
This FCI energy gap vanishes at $V_{\text{imp}}=0.0035$ [Fig.~\ref{Fig:Spectrum}(b)], and there appears a unique gapped ground state, suggesting a zero-temperature phase transition from the FCI state to another insulating state. 
To determine the nature of this state, we evaluate its Hall conductivity by calculating its many-body Chern number,
\begin{align}
	C=\frac1{\pi}\iint\dd{\theta_1}\dd{\theta_2} \Im\braket{\partial_{\theta_1}\psi}{\partial_{\theta_2}\psi},
\end{align}
which can be efficiently calculated by the algorithm devised in Ref.~\cite{Fukui2005}. We find that this insulating state has $C=-1$ as anticipated, corroborating that it is a pinned hole-WC.
  
To study the finite-temperature states, we use the hole entanglement spectrum (HES), the particle-hole conjugate of the particle entanglement spectrum~\cite{Sterdyniak2011}, to characterize the hole-WC and FCI states. 
The HES is defined as the spectrum of $\xi = -\ln(\rho_A)$, where the system is partitioned into subsystem A containing $n_a$ holes and subsystem B containing $(n_h - n_a)$ holes. The reduced density matrix $\rho_A = \mathrm{Tr}_B(\rho)$ is obtained by tracing out subsystem B from the finite-temperature density matrix $\rho=e^{-H/T}/\trace[e^{-H/T}]$. 
The detailed procedure of tracing out $x$ particles/holes is given in the SM~\cite{supplement}. 
Due to the lack of symmetries, we cannot obtain the whole spectrum to calculate the density matrix for this system size. 
Instead, we use the lowest $N_{\text{cutoff}}$ states to approximate the density matrix through
$\rho=Z^{-1}\sum_{i=1}^{N_{\text{cutoff}}} e^{-e_i/T}\dyad{\psi_i}$,
where $e_i$ and $\ket{\psi_i}$ are the eigenvalue and its corresponding eigenstate, and the normalization factor is calculated by $Z=\sum_{i=1}^{N_{\text{cutoff}}}e^{-e_i/T}$. 
We choose $N_{\text{cutoff}}=1000$ throughout the calculations.
The HES reflects the quasiparticle excitations of a state as an entanglement gap in the HES, below which the number of states agrees with the number of quasiparticle excitations.

The quasiparticle excitation of an FCI ground state follows the generalized Pauli exclusion principle~\cite{Regnault2011}, whereas a pinned hole-WC only has $\binom{n_h}{n_a}$ excitations. 
In Fig.~\ref{Fig:Spectrum}(c), we take $n_a=3$ and calculate the HES at finite temperatures for $V_{\text{imp}}=0.0035$. 
Instead of an FCI entanglement gap, there is an entanglement gap with $20$ states below it at low temperatures, which is consistent with the excitations of a pinned hole-WC. 
As the temperature increases, however, the entanglement gap at low temperatures vanishes, giving way to another entanglement gap corresponding to the (1,3)-permissible quasiparticle excitation~\cite{Regnault2011}, which is a hallmark of an FCI state. 
From this HES plot, we can infer that, while the ground state of the system shown in Fig.~\ref{Fig:Spectrum}(b) is a pinned hole-WC at zero temperature, it contains some high-energy FCI-like states.  
Therefore, the system undergoes a crossover from the hole-WC to the FCI phase as the temperature is raised. 
We note that the opening and closing of the entanglement gap at finite temperatures should be regarded as a crossover rather than a genuine phase transition.
It is important to note that the finite-temperature density matrix is a mixed state, incorporating contributions from both the ground state and excited FCI states. Thus, the appearance of a finite FCI entanglement gap at finite temperature indicates that FCI-like states have significant weight in the density matrix, but does not imply that the system is in a pure FCI state.

\begin{figure}[t]
	\center
	\includegraphics[width=0.45\textwidth]{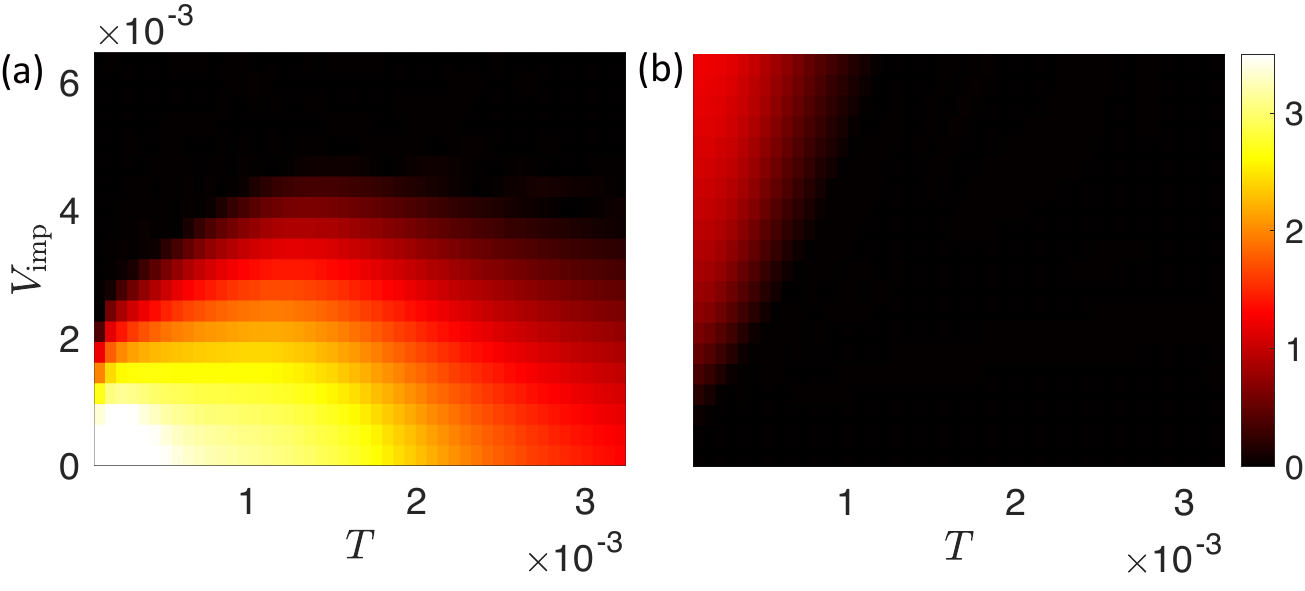}
	\caption{\label{Fig:PES} 
	(a) and (b) are the phase diagrams in terms of the entanglement gap for the FCI and FL, respectively. 
	The other parameters are the same as those used in Fig.~\ref{Fig:density}. 
	}
\end{figure}

In Fig.~\ref{Fig:PES}, we present the phase diagram of the hole-WC and FCI entanglement gap to summarize our findings. 
For $V_{\text{imp}}<0.004$, we observe a crossover from hole-WC to FCI at finite temperatures. 
Moreover, the temperature at which the FCI entanglement gap emerges increases with increasing $V_{\text{imp}}$, consistent with our speculation for $T_e$ in Eq.~\eqref{Eq:T_e}. 
Fig.~\ref{Fig:PES}(a) also shows that there is no obvious crossover to the FCI phase for $V_{\text{imp}}>0.004$, implying that $T_e$ eventually surpasses $T_{\text{FCI}}$ for large $V_\text{imp}$. 
Finally, Fig.~\ref{Fig:PES}(b) shows the entanglement gap for the WC phase. 
The WC entanglement gap appears at large impurity strength and low temperatures, and for $0.002<V_{\text{imp}}<0.004$, there is a crossover between the two entanglement gaps at finite temperatures.

\lettersection{Discussion}
In this Letter, we discuss an impurity-induced mechanism for the thermal crossover in FCI. 
One key aspect of our theory is that when the particle number is slightly less than the exact fractional fillings, the FCI state has a large entropy because of the abundant quasiparticle excitations. 
This can potentially cause a crossover from a low-entropy ground state at low temperatures to a high-entropy FCI state at high temperatures. 
Particularly, we numerically verify the argument in a toy model of FCI, where there is a crossover from a pinned hole-WC to an FCI at $2/3$ filling. 
As the pinned hole-WC has the same Hall conductivity as the completely filled topological band, our theory can explain the crossover from the FQAH phase to the IQAH phase observed in the recent experiments in the PLG~\cite{Lu2025}. 
Our theory for the thermal crossover is also closely related to the experimental observation of the fractional quantum Hall (FQH) state at the $1/7$ filling that appears around $T\approx \SI{100}{mK}$ but disappears for higher or lower temperatures~\cite{Chung2022}.

Another key aspect of our theory is that the thermal crossover only occurs when $T_e < T_\text{FCI}$, suggesting that the thermal crossover requires a delicate balance between the impurity strength and entropy in the system.
For the crossover to be experimentally observable, the system must have an appropriate combination of these factors, such that $T_e < T_\text{FCI}$. 
Due to the strong sample-to-sample variations in impurity strength, we expect this crossover to be observable only in specific samples. 
In cases where $T_e$ is extremely small, the crossover may occur at temperatures far below the current experimental reach. Furthermore, our results suggest that increasing (decreasing) disorder in the system should increase (decrease) the crossover temperature scale from IQAH effect to FQAH effect. 
In the strong disorder limit where $T_e$ exceeds $T_\text{FCI}$, FQAH effect will be completely suppressed. 
It is possible that the $T=0$ phase is always IQAH effect, as all samples inevitably contain some level of impurities. 
These findings could explain why the crossover has been observed only for specific ranges of the displacement field and not in all samples. 
Observing the crossover described in our theory depends on the delicate interplay between impurity strength, entropy, and the accessible temperature range in the experiment.

\lettersection{Acknowledgement}
K.H. and X.L. are supported by the Research Grants Council of Hong Kong (Grants No. CityU 11300421, CityU 11304823, and C7012-21G) and City University of Hong Kong (Project No.~9610428). K.H. is also supported by the Hong Kong PhD Fellowship Scheme. 
S.D.S. is supported by the Laboratory for Physical Sciences through the Condensed Matter Theory Center (CMTC) at the University of Maryland. 
This research was supported in part by grant NSF~PHY-2309135 to the Kavli Institute for Theoretical Physics (KITP).

\bibliography{FCI_dis.bib}

\begin{thebibliography}{30}%
\makeatletter
\providecommand \@ifxundefined [1]{%
 \@ifx{#1\undefined}
}%
\providecommand \@ifnum [1]{%
 \ifnum #1\expandafter \@firstoftwo
 \else \expandafter \@secondoftwo
 \fi
}%
\providecommand \@ifx [1]{%
 \ifx #1\expandafter \@firstoftwo
 \else \expandafter \@secondoftwo
 \fi
}%
\providecommand \natexlab [1]{#1}%
\providecommand \enquote  [1]{``#1''}%
\providecommand \bibnamefont  [1]{#1}%
\providecommand \bibfnamefont [1]{#1}%
\providecommand \citenamefont [1]{#1}%
\providecommand \href@noop [0]{\@secondoftwo}%
\providecommand \href [0]{\begingroup \@sanitize@url \@href}%
\providecommand \@href[1]{\@@startlink{#1}\@@href}%
\providecommand \@@href[1]{\endgroup#1\@@endlink}%
\providecommand \@sanitize@url [0]{\catcode `\\12\catcode `\$12\catcode
  `\&12\catcode `\#12\catcode `\^12\catcode `\_12\catcode `\%12\relax}%
\providecommand \@@startlink[1]{}%
\providecommand \@@endlink[0]{}%
\providecommand \url  [0]{\begingroup\@sanitize@url \@url }%
\providecommand \@url [1]{\endgroup\@href {#1}{\urlprefix }}%
\providecommand \urlprefix  [0]{URL }%
\providecommand \Eprint [0]{\href }%
\providecommand \doibase [0]{https://doi.org/}%
\providecommand \selectlanguage [0]{\@gobble}%
\providecommand \bibinfo  [0]{\@secondoftwo}%
\providecommand \bibfield  [0]{\@secondoftwo}%
\providecommand \translation [1]{[#1]}%
\providecommand \BibitemOpen [0]{}%
\providecommand \bibitemStop [0]{}%
\providecommand \bibitemNoStop [0]{.\EOS\space}%
\providecommand \EOS [0]{\spacefactor3000\relax}%
\providecommand \BibitemShut  [1]{\csname bibitem#1\endcsname}%
\let\auto@bib@innerbib\@empty
\bibitem [{\citenamefont {Tang}\ \emph {et~al.}(2011)\citenamefont {Tang},
  \citenamefont {Mei},\ and\ \citenamefont {Wen}}]{Tang2011}%
  \BibitemOpen
  \bibfield  {author} {\bibinfo {author} {\bibfnamefont {E.}~\bibnamefont
  {Tang}}, \bibinfo {author} {\bibfnamefont {J.-W.}\ \bibnamefont {Mei}},\ and\
  \bibinfo {author} {\bibfnamefont {X.-G.}\ \bibnamefont {Wen}},\ }\bibfield
  {title} {\bibinfo {title} {{High-Temperature Fractional Quantum Hall
  States}},\ }\href {https://doi.org/10.1103/physrevlett.106.236802} {\bibfield
   {journal} {\bibinfo  {journal} {Phys. Rev. Lett.}\ }\textbf {\bibinfo
  {volume} {106}},\ \bibinfo {pages} {236802} (\bibinfo {year}
  {2011})}\BibitemShut {NoStop}%
\bibitem [{\citenamefont {Sun}\ \emph {et~al.}(2011)\citenamefont {Sun},
  \citenamefont {Gu}, \citenamefont {Katsura},\ and\ \citenamefont
  {Das~Sarma}}]{Sun2011}%
  \BibitemOpen
  \bibfield  {author} {\bibinfo {author} {\bibfnamefont {K.}~\bibnamefont
  {Sun}}, \bibinfo {author} {\bibfnamefont {Z.}~\bibnamefont {Gu}}, \bibinfo
  {author} {\bibfnamefont {H.}~\bibnamefont {Katsura}},\ and\ \bibinfo {author}
  {\bibfnamefont {S.}~\bibnamefont {Das~Sarma}},\ }\bibfield  {title} {\bibinfo
  {title} {{Nearly Flatbands with Nontrivial Topology}},\ }\href
  {https://doi.org/10.1103/physrevlett.106.236803} {\bibfield  {journal}
  {\bibinfo  {journal} {Phys. Rev. Lett.}\ }\textbf {\bibinfo {volume} {106}},\
  \bibinfo {pages} {236803} (\bibinfo {year} {2011})}\BibitemShut {NoStop}%
\bibitem [{\citenamefont {Neupert}\ \emph {et~al.}(2011)\citenamefont
  {Neupert}, \citenamefont {Santos}, \citenamefont {Chamon},\ and\
  \citenamefont {Mudry}}]{Neupert2011}%
  \BibitemOpen
  \bibfield  {author} {\bibinfo {author} {\bibfnamefont {T.}~\bibnamefont
  {Neupert}}, \bibinfo {author} {\bibfnamefont {L.}~\bibnamefont {Santos}},
  \bibinfo {author} {\bibfnamefont {C.}~\bibnamefont {Chamon}},\ and\ \bibinfo
  {author} {\bibfnamefont {C.}~\bibnamefont {Mudry}},\ }\bibfield  {title}
  {\bibinfo {title} {{Fractional Quantum {Hall} States at Zero Magnetic
  Field}},\ }\href {https://doi.org/10.1103/physrevlett.106.236804} {\bibfield
  {journal} {\bibinfo  {journal} {Phys. Rev. Lett.}\ }\textbf {\bibinfo
  {volume} {106}},\ \bibinfo {pages} {236804} (\bibinfo {year}
  {2011})}\BibitemShut {NoStop}%
\bibitem [{\citenamefont {Sheng}\ \emph {et~al.}(2011)\citenamefont {Sheng},
  \citenamefont {Gu}, \citenamefont {Sun},\ and\ \citenamefont
  {Sheng}}]{Sheng2011}%
  \BibitemOpen
  \bibfield  {author} {\bibinfo {author} {\bibfnamefont {D.}~\bibnamefont
  {Sheng}}, \bibinfo {author} {\bibfnamefont {Z.-C.}\ \bibnamefont {Gu}},
  \bibinfo {author} {\bibfnamefont {K.}~\bibnamefont {Sun}},\ and\ \bibinfo
  {author} {\bibfnamefont {L.}~\bibnamefont {Sheng}},\ }\bibfield  {title}
  {\bibinfo {title} {{Fractional quantum Hall effect in the absence of Landau
  levels}},\ }\href {https://doi.org/10.1038/ncomms1380} {\bibfield  {journal}
  {\bibinfo  {journal} {Nat. Commun.}\ }\textbf {\bibinfo {volume} {2}},\
  \bibinfo {pages} {389} (\bibinfo {year} {2011})}\BibitemShut {NoStop}%
\bibitem [{\citenamefont {Regnault}\ and\ \citenamefont
  {Bernevig}(2011)}]{Regnault2011}%
  \BibitemOpen
  \bibfield  {author} {\bibinfo {author} {\bibfnamefont {N.}~\bibnamefont
  {Regnault}}\ and\ \bibinfo {author} {\bibfnamefont {B.~A.}\ \bibnamefont
  {Bernevig}},\ }\bibfield  {title} {\bibinfo {title} {{Fractional Chern
  Insulator}},\ }\href {https://doi.org/10.1103/physrevx.1.021014} {\bibfield
  {journal} {\bibinfo  {journal} {Phys. Rev. X}\ }\textbf {\bibinfo {volume}
  {1}},\ \bibinfo {pages} {021014} (\bibinfo {year} {2011})}\BibitemShut
  {NoStop}%
\bibitem [{\citenamefont {Cai}\ \emph {et~al.}(2023)\citenamefont {Cai},
  \citenamefont {Anderson}, \citenamefont {Wang}, \citenamefont {Zhang},
  \citenamefont {Liu}, \citenamefont {Holtzmann}, \citenamefont {Zhang},
  \citenamefont {Fan}, \citenamefont {Taniguchi}, \citenamefont {Watanabe},
  \citenamefont {Ran}, \citenamefont {Cao}, \citenamefont {Fu}, \citenamefont
  {Xiao}, \citenamefont {Yao},\ and\ \citenamefont {Xu}}]{Cai2023}%
  \BibitemOpen
  \bibfield  {author} {\bibinfo {author} {\bibfnamefont {J.}~\bibnamefont
  {Cai}}, \bibinfo {author} {\bibfnamefont {E.}~\bibnamefont {Anderson}},
  \bibinfo {author} {\bibfnamefont {C.}~\bibnamefont {Wang}}, \bibinfo {author}
  {\bibfnamefont {X.}~\bibnamefont {Zhang}}, \bibinfo {author} {\bibfnamefont
  {X.}~\bibnamefont {Liu}}, \bibinfo {author} {\bibfnamefont {W.}~\bibnamefont
  {Holtzmann}}, \bibinfo {author} {\bibfnamefont {Y.}~\bibnamefont {Zhang}},
  \bibinfo {author} {\bibfnamefont {F.}~\bibnamefont {Fan}}, \bibinfo {author}
  {\bibfnamefont {T.}~\bibnamefont {Taniguchi}}, \bibinfo {author}
  {\bibfnamefont {K.}~\bibnamefont {Watanabe}}, \bibinfo {author}
  {\bibfnamefont {Y.}~\bibnamefont {Ran}}, \bibinfo {author} {\bibfnamefont
  {T.}~\bibnamefont {Cao}}, \bibinfo {author} {\bibfnamefont {L.}~\bibnamefont
  {Fu}}, \bibinfo {author} {\bibfnamefont {D.}~\bibnamefont {Xiao}}, \bibinfo
  {author} {\bibfnamefont {W.}~\bibnamefont {Yao}},\ and\ \bibinfo {author}
  {\bibfnamefont {X.}~\bibnamefont {Xu}},\ }\bibfield  {title} {\bibinfo
  {title} {{Signatures of fractional quantum anomalous Hall states in twisted
  MoTe$_2$}},\ }\href {https://doi.org/10.1038/s41586-023-06289-w} {\bibfield
  {journal} {\bibinfo  {journal} {Nature}\ }\textbf {\bibinfo {volume} {622}},\
  \bibinfo {pages} {63} (\bibinfo {year} {2023})}\BibitemShut {NoStop}%
\bibitem [{\citenamefont {Zeng}\ \emph {et~al.}(2023)\citenamefont {Zeng},
  \citenamefont {Xia}, \citenamefont {Kang}, \citenamefont {Zhu}, \citenamefont
  {Knüppel}, \citenamefont {Vaswani}, \citenamefont {Watanabe}, \citenamefont
  {Taniguchi}, \citenamefont {Mak},\ and\ \citenamefont {Shan}}]{Zeng2023}%
  \BibitemOpen
  \bibfield  {author} {\bibinfo {author} {\bibfnamefont {Y.}~\bibnamefont
  {Zeng}}, \bibinfo {author} {\bibfnamefont {Z.}~\bibnamefont {Xia}}, \bibinfo
  {author} {\bibfnamefont {K.}~\bibnamefont {Kang}}, \bibinfo {author}
  {\bibfnamefont {J.}~\bibnamefont {Zhu}}, \bibinfo {author} {\bibfnamefont
  {P.}~\bibnamefont {Knüppel}}, \bibinfo {author} {\bibfnamefont
  {C.}~\bibnamefont {Vaswani}}, \bibinfo {author} {\bibfnamefont
  {K.}~\bibnamefont {Watanabe}}, \bibinfo {author} {\bibfnamefont
  {T.}~\bibnamefont {Taniguchi}}, \bibinfo {author} {\bibfnamefont {K.~F.}\
  \bibnamefont {Mak}},\ and\ \bibinfo {author} {\bibfnamefont {J.}~\bibnamefont
  {Shan}},\ }\bibfield  {title} {\bibinfo {title} {{Thermodynamic evidence of
  fractional Chern insulator in moir\'{e} MoTe$_2$}},\ }\href
  {https://doi.org/10.1038/s41586-023-06452-3} {\bibfield  {journal} {\bibinfo
  {journal} {Nature}\ }\textbf {\bibinfo {volume} {622}},\ \bibinfo {pages}
  {69} (\bibinfo {year} {2023})}\BibitemShut {NoStop}%
\bibitem [{\citenamefont {Park}\ \emph {et~al.}(2023)\citenamefont {Park},
  \citenamefont {Cai}, \citenamefont {Anderson}, \citenamefont {Zhang},
  \citenamefont {Zhu}, \citenamefont {Liu}, \citenamefont {Wang}, \citenamefont
  {Holtzmann}, \citenamefont {Hu}, \citenamefont {Liu}, \citenamefont
  {Taniguchi}, \citenamefont {Watanabe}, \citenamefont {Chu}, \citenamefont
  {Cao}, \citenamefont {Fu}, \citenamefont {Yao}, \citenamefont {Chang},
  \citenamefont {Cobden}, \citenamefont {Xiao},\ and\ \citenamefont
  {Xu}}]{Park2023}%
  \BibitemOpen
  \bibfield  {author} {\bibinfo {author} {\bibfnamefont {H.}~\bibnamefont
  {Park}}, \bibinfo {author} {\bibfnamefont {J.}~\bibnamefont {Cai}}, \bibinfo
  {author} {\bibfnamefont {E.}~\bibnamefont {Anderson}}, \bibinfo {author}
  {\bibfnamefont {Y.}~\bibnamefont {Zhang}}, \bibinfo {author} {\bibfnamefont
  {J.}~\bibnamefont {Zhu}}, \bibinfo {author} {\bibfnamefont {X.}~\bibnamefont
  {Liu}}, \bibinfo {author} {\bibfnamefont {C.}~\bibnamefont {Wang}}, \bibinfo
  {author} {\bibfnamefont {W.}~\bibnamefont {Holtzmann}}, \bibinfo {author}
  {\bibfnamefont {C.}~\bibnamefont {Hu}}, \bibinfo {author} {\bibfnamefont
  {Z.}~\bibnamefont {Liu}}, \bibinfo {author} {\bibfnamefont {T.}~\bibnamefont
  {Taniguchi}}, \bibinfo {author} {\bibfnamefont {K.}~\bibnamefont {Watanabe}},
  \bibinfo {author} {\bibfnamefont {J.-H.}\ \bibnamefont {Chu}}, \bibinfo
  {author} {\bibfnamefont {T.}~\bibnamefont {Cao}}, \bibinfo {author}
  {\bibfnamefont {L.}~\bibnamefont {Fu}}, \bibinfo {author} {\bibfnamefont
  {W.}~\bibnamefont {Yao}}, \bibinfo {author} {\bibfnamefont {C.-Z.}\
  \bibnamefont {Chang}}, \bibinfo {author} {\bibfnamefont {D.}~\bibnamefont
  {Cobden}}, \bibinfo {author} {\bibfnamefont {D.}~\bibnamefont {Xiao}},\ and\
  \bibinfo {author} {\bibfnamefont {X.}~\bibnamefont {Xu}},\ }\bibfield
  {title} {\bibinfo {title} {{Observation of fractionally quantized anomalous
  Hall effect}},\ }\href {https://doi.org/10.1038/s41586-023-06536-0}
  {\bibfield  {journal} {\bibinfo  {journal} {Nature}\ }\textbf {\bibinfo
  {volume} {622}},\ \bibinfo {pages} {74} (\bibinfo {year} {2023})}\BibitemShut
  {NoStop}%
\bibitem [{\citenamefont {Xu}\ \emph {et~al.}(2023)\citenamefont {Xu},
  \citenamefont {Sun}, \citenamefont {Jia}, \citenamefont {Liu}, \citenamefont
  {Xu}, \citenamefont {Li}, \citenamefont {Gu}, \citenamefont {Watanabe},
  \citenamefont {Taniguchi}, \citenamefont {Tong}, \citenamefont {Jia},
  \citenamefont {Shi}, \citenamefont {Jiang}, \citenamefont {Zhang},
  \citenamefont {Liu},\ and\ \citenamefont {Li}}]{Xu2023}%
  \BibitemOpen
  \bibfield  {author} {\bibinfo {author} {\bibfnamefont {F.}~\bibnamefont
  {Xu}}, \bibinfo {author} {\bibfnamefont {Z.}~\bibnamefont {Sun}}, \bibinfo
  {author} {\bibfnamefont {T.}~\bibnamefont {Jia}}, \bibinfo {author}
  {\bibfnamefont {C.}~\bibnamefont {Liu}}, \bibinfo {author} {\bibfnamefont
  {C.}~\bibnamefont {Xu}}, \bibinfo {author} {\bibfnamefont {C.}~\bibnamefont
  {Li}}, \bibinfo {author} {\bibfnamefont {Y.}~\bibnamefont {Gu}}, \bibinfo
  {author} {\bibfnamefont {K.}~\bibnamefont {Watanabe}}, \bibinfo {author}
  {\bibfnamefont {T.}~\bibnamefont {Taniguchi}}, \bibinfo {author}
  {\bibfnamefont {B.}~\bibnamefont {Tong}}, \bibinfo {author} {\bibfnamefont
  {J.}~\bibnamefont {Jia}}, \bibinfo {author} {\bibfnamefont {Z.}~\bibnamefont
  {Shi}}, \bibinfo {author} {\bibfnamefont {S.}~\bibnamefont {Jiang}}, \bibinfo
  {author} {\bibfnamefont {Y.}~\bibnamefont {Zhang}}, \bibinfo {author}
  {\bibfnamefont {X.}~\bibnamefont {Liu}},\ and\ \bibinfo {author}
  {\bibfnamefont {T.}~\bibnamefont {Li}},\ }\bibfield  {title} {\bibinfo
  {title} {{Observation of Integer and Fractional Quantum Anomalous Hall
  Effects in Twisted Bilayer MoTe$_2$}},\ }\href
  {https://doi.org/10.1103/physrevx.13.031037} {\bibfield  {journal} {\bibinfo
  {journal} {Phys. Rev. X}\ }\textbf {\bibinfo {volume} {13}},\ \bibinfo
  {pages} {031037} (\bibinfo {year} {2023})}\BibitemShut {NoStop}%
\bibitem [{\citenamefont {Lu}\ \emph {et~al.}(2024{\natexlab{a}})\citenamefont
  {Lu}, \citenamefont {Han}, \citenamefont {Yao}, \citenamefont {Reddy},
  \citenamefont {Yang}, \citenamefont {Seo}, \citenamefont {Watanabe},
  \citenamefont {Taniguchi}, \citenamefont {Fu},\ and\ \citenamefont
  {Ju}}]{Lu2024}%
  \BibitemOpen
  \bibfield  {author} {\bibinfo {author} {\bibfnamefont {Z.}~\bibnamefont
  {Lu}}, \bibinfo {author} {\bibfnamefont {T.}~\bibnamefont {Han}}, \bibinfo
  {author} {\bibfnamefont {Y.}~\bibnamefont {Yao}}, \bibinfo {author}
  {\bibfnamefont {A.~P.}\ \bibnamefont {Reddy}}, \bibinfo {author}
  {\bibfnamefont {J.}~\bibnamefont {Yang}}, \bibinfo {author} {\bibfnamefont
  {J.}~\bibnamefont {Seo}}, \bibinfo {author} {\bibfnamefont {K.}~\bibnamefont
  {Watanabe}}, \bibinfo {author} {\bibfnamefont {T.}~\bibnamefont {Taniguchi}},
  \bibinfo {author} {\bibfnamefont {L.}~\bibnamefont {Fu}},\ and\ \bibinfo
  {author} {\bibfnamefont {L.}~\bibnamefont {Ju}},\ }\bibfield  {title}
  {\bibinfo {title} {{Fractional quantum anomalous Hall effect in multilayer
  graphene}},\ }\href {https://doi.org/10.1038/s41586-023-07010-7} {\bibfield
  {journal} {\bibinfo  {journal} {Nature}\ }\textbf {\bibinfo {volume} {626}},\
  \bibinfo {pages} {759} (\bibinfo {year} {2024}{\natexlab{a}})}\BibitemShut
  {NoStop}%
\bibitem [{\citenamefont {Xie}\ \emph {et~al.}(2025)\citenamefont {Xie},
  \citenamefont {Huo}, \citenamefont {Lu}, \citenamefont {Feng}, \citenamefont
  {Zhang}, \citenamefont {Wang}, \citenamefont {Yang}, \citenamefont
  {Watanabe}, \citenamefont {Taniguchi}, \citenamefont {Liu}, \citenamefont
  {Song}, \citenamefont {Xie}, \citenamefont {Liu},\ and\ \citenamefont
  {Lu}}]{XieJian2025}%
  \BibitemOpen
  \bibfield  {author} {\bibinfo {author} {\bibfnamefont {J.}~\bibnamefont
  {Xie}}, \bibinfo {author} {\bibfnamefont {Z.}~\bibnamefont {Huo}}, \bibinfo
  {author} {\bibfnamefont {X.}~\bibnamefont {Lu}}, \bibinfo {author}
  {\bibfnamefont {Z.}~\bibnamefont {Feng}}, \bibinfo {author} {\bibfnamefont
  {Z.}~\bibnamefont {Zhang}}, \bibinfo {author} {\bibfnamefont
  {W.}~\bibnamefont {Wang}}, \bibinfo {author} {\bibfnamefont {Q.}~\bibnamefont
  {Yang}}, \bibinfo {author} {\bibfnamefont {K.}~\bibnamefont {Watanabe}},
  \bibinfo {author} {\bibfnamefont {T.}~\bibnamefont {Taniguchi}}, \bibinfo
  {author} {\bibfnamefont {K.}~\bibnamefont {Liu}}, \bibinfo {author}
  {\bibfnamefont {Z.}~\bibnamefont {Song}}, \bibinfo {author} {\bibfnamefont
  {X.~C.}\ \bibnamefont {Xie}}, \bibinfo {author} {\bibfnamefont
  {J.}~\bibnamefont {Liu}},\ and\ \bibinfo {author} {\bibfnamefont
  {X.}~\bibnamefont {Lu}},\ }\bibfield  {title} {\bibinfo {title} {Tunable
  fractional chern insulators in rhombohedral graphene superlattices},\ }\href
  {https://doi.org/10.1038/s41563-025-02225-7} {\bibfield  {journal} {\bibinfo
  {journal} {Nat. Mater.}\ }\textbf {\bibinfo {volume} {24}},\ \bibinfo {pages}
  {1042} (\bibinfo {year} {2025})}\BibitemShut {NoStop}%
\bibitem [{\citenamefont {Waters}\ \emph {et~al.}(2025)\citenamefont {Waters},
  \citenamefont {Okounkova}, \citenamefont {Su}, \citenamefont {Zhou},
  \citenamefont {Yao}, \citenamefont {Watanabe}, \citenamefont {Taniguchi},
  \citenamefont {Xu}, \citenamefont {Zhang}, \citenamefont {Folk},\ and\
  \citenamefont {Yankowitz}}]{Waters2025}%
  \BibitemOpen
  \bibfield  {author} {\bibinfo {author} {\bibfnamefont {D.}~\bibnamefont
  {Waters}}, \bibinfo {author} {\bibfnamefont {A.}~\bibnamefont {Okounkova}},
  \bibinfo {author} {\bibfnamefont {R.}~\bibnamefont {Su}}, \bibinfo {author}
  {\bibfnamefont {B.}~\bibnamefont {Zhou}}, \bibinfo {author} {\bibfnamefont
  {J.}~\bibnamefont {Yao}}, \bibinfo {author} {\bibfnamefont {K.}~\bibnamefont
  {Watanabe}}, \bibinfo {author} {\bibfnamefont {T.}~\bibnamefont {Taniguchi}},
  \bibinfo {author} {\bibfnamefont {X.}~\bibnamefont {Xu}}, \bibinfo {author}
  {\bibfnamefont {Y.-H.}\ \bibnamefont {Zhang}}, \bibinfo {author}
  {\bibfnamefont {J.}~\bibnamefont {Folk}},\ and\ \bibinfo {author}
  {\bibfnamefont {M.}~\bibnamefont {Yankowitz}},\ }\bibfield  {title} {\bibinfo
  {title} {Chern insulators at integer and fractional filling in moiré
  pentalayer graphene},\ }\href {https://doi.org/10.1103/physrevx.15.011045}
  {\bibfield  {journal} {\bibinfo  {journal} {Phys. Rev. X}\ }\textbf {\bibinfo
  {volume} {15}},\ \bibinfo {pages} {011045} (\bibinfo {year}
  {2025})}\BibitemShut {NoStop}%
\bibitem [{\citenamefont {Lu}\ \emph {et~al.}(2025)\citenamefont {Lu},
  \citenamefont {Han}, \citenamefont {Yao}, \citenamefont {Hadjri},
  \citenamefont {Yang}, \citenamefont {Seo}, \citenamefont {Shi}, \citenamefont
  {Ye}, \citenamefont {Watanabe}, \citenamefont {Taniguchi},\ and\
  \citenamefont {Ju}}]{Lu2025}%
  \BibitemOpen
  \bibfield  {author} {\bibinfo {author} {\bibfnamefont {Z.}~\bibnamefont
  {Lu}}, \bibinfo {author} {\bibfnamefont {T.}~\bibnamefont {Han}}, \bibinfo
  {author} {\bibfnamefont {Y.}~\bibnamefont {Yao}}, \bibinfo {author}
  {\bibfnamefont {Z.}~\bibnamefont {Hadjri}}, \bibinfo {author} {\bibfnamefont
  {J.}~\bibnamefont {Yang}}, \bibinfo {author} {\bibfnamefont {J.}~\bibnamefont
  {Seo}}, \bibinfo {author} {\bibfnamefont {L.}~\bibnamefont {Shi}}, \bibinfo
  {author} {\bibfnamefont {S.}~\bibnamefont {Ye}}, \bibinfo {author}
  {\bibfnamefont {K.}~\bibnamefont {Watanabe}}, \bibinfo {author}
  {\bibfnamefont {T.}~\bibnamefont {Taniguchi}},\ and\ \bibinfo {author}
  {\bibfnamefont {L.}~\bibnamefont {Ju}},\ }\bibfield  {title} {\bibinfo
  {title} {Extended quantum anomalous hall states in graphene/hbn moiré
  superlattices},\ }\href {https://doi.org/10.1038/s41586-024-08470-1}
  {\bibfield  {journal} {\bibinfo  {journal} {Nature}\ }\textbf {\bibinfo
  {volume} {637}},\ \bibinfo {pages} {1090} (\bibinfo {year}
  {2025})}\BibitemShut {NoStop}%
\bibitem [{\citenamefont {Choi}\ \emph {et~al.}(2025)\citenamefont {Choi},
  \citenamefont {Choi}, \citenamefont {Valentini}, \citenamefont {Patterson},
  \citenamefont {Holleis}, \citenamefont {Sheekey}, \citenamefont {Stoyanov},
  \citenamefont {Cheng}, \citenamefont {Taniguchi}, \citenamefont {Watanabe},\
  and\ \citenamefont {Young}}]{Choi2025}%
  \BibitemOpen
  \bibfield  {author} {\bibinfo {author} {\bibfnamefont {Y.}~\bibnamefont
  {Choi}}, \bibinfo {author} {\bibfnamefont {Y.}~\bibnamefont {Choi}}, \bibinfo
  {author} {\bibfnamefont {M.}~\bibnamefont {Valentini}}, \bibinfo {author}
  {\bibfnamefont {C.~L.}\ \bibnamefont {Patterson}}, \bibinfo {author}
  {\bibfnamefont {L.~F.~W.}\ \bibnamefont {Holleis}}, \bibinfo {author}
  {\bibfnamefont {O.~I.}\ \bibnamefont {Sheekey}}, \bibinfo {author}
  {\bibfnamefont {H.}~\bibnamefont {Stoyanov}}, \bibinfo {author}
  {\bibfnamefont {X.}~\bibnamefont {Cheng}}, \bibinfo {author} {\bibfnamefont
  {T.}~\bibnamefont {Taniguchi}}, \bibinfo {author} {\bibfnamefont
  {K.}~\bibnamefont {Watanabe}},\ and\ \bibinfo {author} {\bibfnamefont
  {A.~F.}\ \bibnamefont {Young}},\ }\bibfield  {title} {\bibinfo {title}
  {Superconductivity and quantized anomalous hall effect in rhombohedral
  graphene},\ }\href {https://doi.org/10.1038/s41586-025-08621-y} {\bibfield
  {journal} {\bibinfo  {journal} {Nature}\ }\textbf {\bibinfo {volume} {639}},\
  \bibinfo {pages} {342} (\bibinfo {year} {2025})}\BibitemShut {NoStop}%
\bibitem [{\citenamefont {Lu}\ \emph {et~al.}(2024{\natexlab{b}})\citenamefont
  {Lu}, \citenamefont {Chen}, \citenamefont {Wu}, \citenamefont {Sun},\ and\
  \citenamefont {Meng}}]{LuHongYu2024}%
  \BibitemOpen
  \bibfield  {author} {\bibinfo {author} {\bibfnamefont {H.}~\bibnamefont
  {Lu}}, \bibinfo {author} {\bibfnamefont {B.-B.}\ \bibnamefont {Chen}},
  \bibinfo {author} {\bibfnamefont {H.-Q.}\ \bibnamefont {Wu}}, \bibinfo
  {author} {\bibfnamefont {K.}~\bibnamefont {Sun}},\ and\ \bibinfo {author}
  {\bibfnamefont {Z.~Y.}\ \bibnamefont {Meng}},\ }\bibfield  {title} {\bibinfo
  {title} {{Thermodynamic Response and Neutral Excitations in Integer and
  Fractional Quantum Anomalous Hall States Emerging from Correlated Flat
  Bands}},\ }\href {https://doi.org/10.1103/physrevlett.132.236502} {\bibfield
  {journal} {\bibinfo  {journal} {Phys. Rev. Lett.}\ }\textbf {\bibinfo
  {volume} {132}},\ \bibinfo {pages} {236502} (\bibinfo {year}
  {2024}{\natexlab{b}})}\BibitemShut {NoStop}%
\bibitem [{\citenamefont {Das~Sarma}\ and\ \citenamefont
  {Xie}(2024)}]{DasSarma2024}%
  \BibitemOpen
  \bibfield  {author} {\bibinfo {author} {\bibfnamefont {S.}~\bibnamefont
  {Das~Sarma}}\ and\ \bibinfo {author} {\bibfnamefont {M.}~\bibnamefont
  {Xie}},\ }\bibfield  {title} {\bibinfo {title} {Thermal crossover from a
  chern insulator to a fractional chern insulator in pentalayer graphene},\
  }\href {https://doi.org/10.1103/physrevb.110.155148} {\bibfield  {journal}
  {\bibinfo  {journal} {Phys. Rev. B}\ }\textbf {\bibinfo {volume} {110}},\
  \bibinfo {pages} {155148} (\bibinfo {year} {2024})}\BibitemShut {NoStop}%
\bibitem [{\citenamefont {Tarnopolsky}\ \emph {et~al.}(2019)\citenamefont
  {Tarnopolsky}, \citenamefont {Kruchkov},\ and\ \citenamefont
  {Vishwanath}}]{Tarnopolsky2019}%
  \BibitemOpen
  \bibfield  {author} {\bibinfo {author} {\bibfnamefont {G.}~\bibnamefont
  {Tarnopolsky}}, \bibinfo {author} {\bibfnamefont {A.~J.}\ \bibnamefont
  {Kruchkov}},\ and\ \bibinfo {author} {\bibfnamefont {A.}~\bibnamefont
  {Vishwanath}},\ }\bibfield  {title} {\bibinfo {title} {{Origin of Magic
  Angles in Twisted Bilayer Graphene}},\ }\href
  {https://doi.org/10.1103/physrevlett.122.106405} {\bibfield  {journal}
  {\bibinfo  {journal} {Phys. Rev. Lett.}\ }\textbf {\bibinfo {volume} {122}},\
  \bibinfo {pages} {106405} (\bibinfo {year} {2019})}\BibitemShut {NoStop}%
\bibitem [{\citenamefont {Wang}\ \emph {et~al.}(2021)\citenamefont {Wang},
  \citenamefont {Cano}, \citenamefont {Millis}, \citenamefont {Liu},\ and\
  \citenamefont {Yang}}]{Wang2021}%
  \BibitemOpen
  \bibfield  {author} {\bibinfo {author} {\bibfnamefont {J.}~\bibnamefont
  {Wang}}, \bibinfo {author} {\bibfnamefont {J.}~\bibnamefont {Cano}}, \bibinfo
  {author} {\bibfnamefont {A.~J.}\ \bibnamefont {Millis}}, \bibinfo {author}
  {\bibfnamefont {Z.}~\bibnamefont {Liu}},\ and\ \bibinfo {author}
  {\bibfnamefont {B.}~\bibnamefont {Yang}},\ }\bibfield  {title} {\bibinfo
  {title} {{Exact Landau Level Description of Geometry and Interaction in a
  Flatband}},\ }\href {https://doi.org/10.1103/physrevlett.127.246403}
  {\bibfield  {journal} {\bibinfo  {journal} {Phys. Rev. Lett.}\ }\textbf
  {\bibinfo {volume} {127}},\ \bibinfo {pages} {246403} (\bibinfo {year}
  {2021})}\BibitemShut {NoStop}%
\bibitem [{\citenamefont {Abouelkomsan}\ \emph {et~al.}(2020)\citenamefont
  {Abouelkomsan}, \citenamefont {Liu},\ and\ \citenamefont
  {Bergholtz}}]{Abouelkomsan2020}%
  \BibitemOpen
  \bibfield  {author} {\bibinfo {author} {\bibfnamefont {A.}~\bibnamefont
  {Abouelkomsan}}, \bibinfo {author} {\bibfnamefont {Z.}~\bibnamefont {Liu}},\
  and\ \bibinfo {author} {\bibfnamefont {E.~J.}\ \bibnamefont {Bergholtz}},\
  }\bibfield  {title} {\bibinfo {title} {{Particle-Hole Duality, Emergent Fermi
  Liquids, and Fractional Chern Insulators in Moir\'e Flatbands}},\ }\href
  {https://doi.org/10.1103/physrevlett.124.106803} {\bibfield  {journal}
  {\bibinfo  {journal} {Phys. Rev. Lett.}\ }\textbf {\bibinfo {volume} {124}},\
  \bibinfo {pages} {106803} (\bibinfo {year} {2020})}\BibitemShut {NoStop}%
\bibitem [{sup()}]{supplement}%
  \BibitemOpen
  \href@noop {} {}\bibinfo {note} {See the Supplementary Material for the
  numerical results in a larger system with $6\times 4$ unit
  cells.}\BibitemShut {Stop}%
\bibitem [{\citenamefont {Platzman}\ and\ \citenamefont
  {Price}(1993)}]{Platzman1993}%
  \BibitemOpen
  \bibfield  {author} {\bibinfo {author} {\bibfnamefont {P.~M.}\ \bibnamefont
  {Platzman}}\ and\ \bibinfo {author} {\bibfnamefont {R.}~\bibnamefont
  {Price}},\ }\bibfield  {title} {\bibinfo {title} {{Quantum freezing of the
  fractional quantum Hall liquid}},\ }\href
  {https://doi.org/10.1103/physrevlett.70.3487} {\bibfield  {journal} {\bibinfo
   {journal} {Phys. Rev. Lett.}\ }\textbf {\bibinfo {volume} {70}},\ \bibinfo
  {pages} {3487} (\bibinfo {year} {1993})}\BibitemShut {NoStop}%
\bibitem [{\citenamefont {Dong}\ \emph {et~al.}(2023)\citenamefont {Dong},
  \citenamefont {Wang}, \citenamefont {Ledwith}, \citenamefont {Vishwanath},\
  and\ \citenamefont {Parker}}]{Dong2023c}%
  \BibitemOpen
  \bibfield  {author} {\bibinfo {author} {\bibfnamefont {J.}~\bibnamefont
  {Dong}}, \bibinfo {author} {\bibfnamefont {J.}~\bibnamefont {Wang}}, \bibinfo
  {author} {\bibfnamefont {P.~J.}\ \bibnamefont {Ledwith}}, \bibinfo {author}
  {\bibfnamefont {A.}~\bibnamefont {Vishwanath}},\ and\ \bibinfo {author}
  {\bibfnamefont {D.~E.}\ \bibnamefont {Parker}},\ }\bibfield  {title}
  {\bibinfo {title} {{Composite Fermi Liquid at Zero Magnetic Field in Twisted
  MoTe$_2$}},\ }\href {https://doi.org/10.1103/physrevlett.131.136502}
  {\bibfield  {journal} {\bibinfo  {journal} {Phys. Rev. Lett.}\ }\textbf
  {\bibinfo {volume} {131}},\ \bibinfo {pages} {136502} (\bibinfo {year}
  {2023})}\BibitemShut {NoStop}%
\bibitem [{\citenamefont {Nuckolls}\ \emph {et~al.}(2020)\citenamefont
  {Nuckolls}, \citenamefont {Oh}, \citenamefont {Wong}, \citenamefont {Lian},
  \citenamefont {Watanabe}, \citenamefont {Taniguchi}, \citenamefont
  {Bernevig},\ and\ \citenamefont {Yazdani}}]{Nuckolls2020}%
  \BibitemOpen
  \bibfield  {author} {\bibinfo {author} {\bibfnamefont {K.~P.}\ \bibnamefont
  {Nuckolls}}, \bibinfo {author} {\bibfnamefont {M.}~\bibnamefont {Oh}},
  \bibinfo {author} {\bibfnamefont {D.}~\bibnamefont {Wong}}, \bibinfo {author}
  {\bibfnamefont {B.}~\bibnamefont {Lian}}, \bibinfo {author} {\bibfnamefont
  {K.}~\bibnamefont {Watanabe}}, \bibinfo {author} {\bibfnamefont
  {T.}~\bibnamefont {Taniguchi}}, \bibinfo {author} {\bibfnamefont {B.~A.}\
  \bibnamefont {Bernevig}},\ and\ \bibinfo {author} {\bibfnamefont
  {A.}~\bibnamefont {Yazdani}},\ }\bibfield  {title} {\bibinfo {title}
  {{Strongly correlated Chern insulators in magic-angle twisted bilayer
  graphene}},\ }\href {https://doi.org/10.1038/s41586-020-3028-8} {\bibfield
  {journal} {\bibinfo  {journal} {Nature}\ }\textbf {\bibinfo {volume} {588}},\
  \bibinfo {pages} {610} (\bibinfo {year} {2020})}\BibitemShut {NoStop}%
\bibitem [{\citenamefont {Lian}\ \emph {et~al.}(2021)\citenamefont {Lian},
  \citenamefont {Song}, \citenamefont {Regnault}, \citenamefont {Efetov},
  \citenamefont {Yazdani},\ and\ \citenamefont {Bernevig}}]{Lian2021}%
  \BibitemOpen
  \bibfield  {author} {\bibinfo {author} {\bibfnamefont {B.}~\bibnamefont
  {Lian}}, \bibinfo {author} {\bibfnamefont {Z.-D.}\ \bibnamefont {Song}},
  \bibinfo {author} {\bibfnamefont {N.}~\bibnamefont {Regnault}}, \bibinfo
  {author} {\bibfnamefont {D.~K.}\ \bibnamefont {Efetov}}, \bibinfo {author}
  {\bibfnamefont {A.}~\bibnamefont {Yazdani}},\ and\ \bibinfo {author}
  {\bibfnamefont {B.~A.}\ \bibnamefont {Bernevig}},\ }\bibfield  {title}
  {\bibinfo {title} {{Twisted bilayer graphene. IV. Exact insulator ground
  states and phase diagram}},\ }\href
  {https://doi.org/10.1103/physrevb.103.205414} {\bibfield  {journal} {\bibinfo
   {journal} {Phys. Rev. B}\ }\textbf {\bibinfo {volume} {103}},\ \bibinfo
  {pages} {205414} (\bibinfo {year} {2021})}\BibitemShut {NoStop}%
\bibitem [{\citenamefont {Bernevig}\ \emph {et~al.}(2021)\citenamefont
  {Bernevig}, \citenamefont {Song}, \citenamefont {Regnault},\ and\
  \citenamefont {Lian}}]{Bernevig2021}%
  \BibitemOpen
  \bibfield  {author} {\bibinfo {author} {\bibfnamefont {B.~A.}\ \bibnamefont
  {Bernevig}}, \bibinfo {author} {\bibfnamefont {Z.-D.}\ \bibnamefont {Song}},
  \bibinfo {author} {\bibfnamefont {N.}~\bibnamefont {Regnault}},\ and\
  \bibinfo {author} {\bibfnamefont {B.}~\bibnamefont {Lian}},\ }\bibfield
  {title} {\bibinfo {title} {{Twisted bilayer graphene. III. Interacting
  Hamiltonian and exact symmetries}},\ }\href
  {https://doi.org/10.1103/physrevb.103.205413} {\bibfield  {journal} {\bibinfo
   {journal} {Phys. Rev. B}\ }\textbf {\bibinfo {volume} {103}},\ \bibinfo
  {pages} {205413} (\bibinfo {year} {2021})}\BibitemShut {NoStop}%
\bibitem [{\citenamefont {Christos}\ \emph {et~al.}(2022)\citenamefont
  {Christos}, \citenamefont {Sachdev},\ and\ \citenamefont
  {Scheurer}}]{Christos2022}%
  \BibitemOpen
  \bibfield  {author} {\bibinfo {author} {\bibfnamefont {M.}~\bibnamefont
  {Christos}}, \bibinfo {author} {\bibfnamefont {S.}~\bibnamefont {Sachdev}},\
  and\ \bibinfo {author} {\bibfnamefont {M.~S.}\ \bibnamefont {Scheurer}},\
  }\bibfield  {title} {\bibinfo {title} {{Correlated Insulators, Semimetals,
  and Superconductivity in Twisted Trilayer Graphene}},\ }\href
  {https://doi.org/10.1103/physrevx.12.021018} {\bibfield  {journal} {\bibinfo
  {journal} {Phys. Rev. X}\ }\textbf {\bibinfo {volume} {12}},\ \bibinfo
  {pages} {021018} (\bibinfo {year} {2022})}\BibitemShut {NoStop}%
\bibitem [{\citenamefont {Kwan}\ \emph {et~al.}(2023)\citenamefont {Kwan},
  \citenamefont {Yu}, \citenamefont {Herzog-Arbeitman}, \citenamefont {Efetov},
  \citenamefont {Regnault},\ and\ \citenamefont {Bernevig}}]{Kwan2023}%
  \BibitemOpen
  \bibfield  {author} {\bibinfo {author} {\bibfnamefont {Y.~H.}\ \bibnamefont
  {Kwan}}, \bibinfo {author} {\bibfnamefont {J.}~\bibnamefont {Yu}}, \bibinfo
  {author} {\bibfnamefont {J.}~\bibnamefont {Herzog-Arbeitman}}, \bibinfo
  {author} {\bibfnamefont {D.~K.}\ \bibnamefont {Efetov}}, \bibinfo {author}
  {\bibfnamefont {N.}~\bibnamefont {Regnault}},\ and\ \bibinfo {author}
  {\bibfnamefont {B.~A.}\ \bibnamefont {Bernevig}},\ }\href
  {https://doi.org/10.48550/ARXIV.2312.11617} {\bibinfo {title} {{Moir\'{e}
  Fractional Chern Insulators III: Hartree-Fock Phase Diagram, Magic Angle
  Regime for Chern Insulator States, the Role of the Moir\'{e} Potential and
  Goldstone Gaps in Rhombohedral Graphene Superlattices}}} (\bibinfo {year}
  {2023}),\ \Eprint {https://arxiv.org/abs/2312.11617} {arXiv:2312.11617
  [cond-mat.str-el]} \BibitemShut {NoStop}%
\bibitem [{\citenamefont {Fukui}\ \emph {et~al.}(2005)\citenamefont {Fukui},
  \citenamefont {Hatsugai},\ and\ \citenamefont {Suzuki}}]{Fukui2005}%
  \BibitemOpen
  \bibfield  {author} {\bibinfo {author} {\bibfnamefont {T.}~\bibnamefont
  {Fukui}}, \bibinfo {author} {\bibfnamefont {Y.}~\bibnamefont {Hatsugai}},\
  and\ \bibinfo {author} {\bibfnamefont {H.}~\bibnamefont {Suzuki}},\
  }\bibfield  {title} {\bibinfo {title} {{Chern Numbers in Discretized
  Brillouin Zone: Efficient Method of Computing (Spin) Hall Conductances}},\
  }\href {https://doi.org/10.1143/jpsj.74.1674} {\bibfield  {journal} {\bibinfo
   {journal} {J. Phys. Soc. Japan}\ }\textbf {\bibinfo {volume} {74}},\
  \bibinfo {pages} {1674} (\bibinfo {year} {2005})}\BibitemShut {NoStop}%
\bibitem [{\citenamefont {Sterdyniak}\ \emph {et~al.}(2011)\citenamefont
  {Sterdyniak}, \citenamefont {Regnault},\ and\ \citenamefont
  {Bernevig}}]{Sterdyniak2011}%
  \BibitemOpen
  \bibfield  {author} {\bibinfo {author} {\bibfnamefont {A.}~\bibnamefont
  {Sterdyniak}}, \bibinfo {author} {\bibfnamefont {N.}~\bibnamefont
  {Regnault}},\ and\ \bibinfo {author} {\bibfnamefont {B.~A.}\ \bibnamefont
  {Bernevig}},\ }\bibfield  {title} {\bibinfo {title} {{Extracting Excitations
  from Model State Entanglement}},\ }\href
  {https://doi.org/10.1103/physrevlett.106.100405} {\bibfield  {journal}
  {\bibinfo  {journal} {Phys. Rev. Lett.}\ }\textbf {\bibinfo {volume} {106}},\
  \bibinfo {pages} {100405} (\bibinfo {year} {2011})}\BibitemShut {NoStop}%
\bibitem [{\citenamefont {Chung}\ \emph {et~al.}(2022)\citenamefont {Chung},
  \citenamefont {Graf}, \citenamefont {Engel}, \citenamefont {Rosales},
  \citenamefont {Madathil}, \citenamefont {Baldwin}, \citenamefont {West},
  \citenamefont {Pfeiffer},\ and\ \citenamefont {Shayegan}}]{Chung2022}%
  \BibitemOpen
  \bibfield  {author} {\bibinfo {author} {\bibfnamefont {Y.~J.}\ \bibnamefont
  {Chung}}, \bibinfo {author} {\bibfnamefont {D.}~\bibnamefont {Graf}},
  \bibinfo {author} {\bibfnamefont {L.}~\bibnamefont {Engel}}, \bibinfo
  {author} {\bibfnamefont {K.~V.}\ \bibnamefont {Rosales}}, \bibinfo {author}
  {\bibfnamefont {P.}~\bibnamefont {Madathil}}, \bibinfo {author}
  {\bibfnamefont {K.}~\bibnamefont {Baldwin}}, \bibinfo {author} {\bibfnamefont
  {K.}~\bibnamefont {West}}, \bibinfo {author} {\bibfnamefont {L.}~\bibnamefont
  {Pfeiffer}},\ and\ \bibinfo {author} {\bibfnamefont {M.}~\bibnamefont
  {Shayegan}},\ }\bibfield  {title} {\bibinfo {title} {{Correlated States of 2D
  Electrons near the Landau Level Filling $\nu=1/7$}},\ }\href
  {https://doi.org/10.1103/physrevlett.128.026802} {\bibfield  {journal}
  {\bibinfo  {journal} {Phys. Rev. Lett.}\ }\textbf {\bibinfo {volume} {128}},\
  \bibinfo {pages} {026802} (\bibinfo {year} {2022})}\BibitemShut {NoStop}%
\end{thebibliography}%


\begin{thebibliography}{2}%
\makeatletter
\providecommand \@ifxundefined [1]{%
 \@ifx{#1\undefined}
}%
\providecommand \@ifnum [1]{%
 \ifnum #1\expandafter \@firstoftwo
 \else \expandafter \@secondoftwo
 \fi
}%
\providecommand \@ifx [1]{%
 \ifx #1\expandafter \@firstoftwo
 \else \expandafter \@secondoftwo
 \fi
}%
\providecommand \natexlab [1]{#1}%
\providecommand \enquote  [1]{``#1''}%
\providecommand \bibnamefont  [1]{#1}%
\providecommand \bibfnamefont [1]{#1}%
\providecommand \citenamefont [1]{#1}%
\providecommand \href@noop [0]{\@secondoftwo}%
\providecommand \href [0]{\begingroup \@sanitize@url \@href}%
\providecommand \@href[1]{\@@startlink{#1}\@@href}%
\providecommand \@@href[1]{\endgroup#1\@@endlink}%
\providecommand \@sanitize@url [0]{\catcode `\\12\catcode `\$12\catcode `\&12\catcode `\#12\catcode `\^12\catcode `\_12\catcode `\%12\relax}%
\providecommand \@@startlink[1]{}%
\providecommand \@@endlink[0]{}%
\providecommand \url  [0]{\begingroup\@sanitize@url \@url }%
\providecommand \@url [1]{\endgroup\@href {#1}{\urlprefix }}%
\providecommand \urlprefix  [0]{URL }%
\providecommand \Eprint [0]{\href }%
\providecommand \doibase [0]{https://doi.org/}%
\providecommand \selectlanguage [0]{\@gobble}%
\providecommand \bibinfo  [0]{\@secondoftwo}%
\providecommand \bibfield  [0]{\@secondoftwo}%
\providecommand \translation [1]{[#1]}%
\providecommand \BibitemOpen [0]{}%
\providecommand \bibitemStop [0]{}%
\providecommand \bibitemNoStop [0]{.\EOS\space}%
\providecommand \EOS [0]{\spacefactor3000\relax}%
\providecommand \BibitemShut  [1]{\csname bibitem#1\endcsname}%
\let\auto@bib@innerbib\@empty
\bibitem [{\citenamefont {Platzman}\ and\ \citenamefont {Price}(1993)}]{Platzman1993}%
  \BibitemOpen
  \bibfield  {author} {\bibinfo {author} {\bibfnamefont {P.~M.}\ \bibnamefont {Platzman}}\ and\ \bibinfo {author} {\bibfnamefont {R.}~\bibnamefont {Price}},\ }\href {https://doi.org/10.1103/physrevlett.70.3487} {\bibfield  {journal} {\bibinfo  {journal} {Phys. Rev. Lett.}\ }\textbf {\bibinfo {volume} {70}},\ \bibinfo {pages} {3487} (\bibinfo {year} {1993})}\BibitemShut {NoStop}%
\bibitem [{\citenamefont {Dong}\ \emph {et~al.}(2023)\citenamefont {Dong}, \citenamefont {Wang}, \citenamefont {Ledwith}, \citenamefont {Vishwanath},\ and\ \citenamefont {Parker}}]{Dong2023c}%
  \BibitemOpen
  \bibfield  {author} {\bibinfo {author} {\bibfnamefont {J.}~\bibnamefont {Dong}}, \bibinfo {author} {\bibfnamefont {J.}~\bibnamefont {Wang}}, \bibinfo {author} {\bibfnamefont {P.~J.}\ \bibnamefont {Ledwith}}, \bibinfo {author} {\bibfnamefont {A.}~\bibnamefont {Vishwanath}},\ and\ \bibinfo {author} {\bibfnamefont {D.~E.}\ \bibnamefont {Parker}},\ }\href {https://doi.org/10.1103/physrevlett.131.136502} {\bibfield  {journal} {\bibinfo  {journal} {Phys. Rev. Lett.}\ }\textbf {\bibinfo {volume} {131}},\ \bibinfo {pages} {136502} (\bibinfo {year} {2023})}\BibitemShut {NoStop}%
\end{thebibliography}%

\end{document}


\global\long\def\id{\mathbbm{1}}
\global\long\def\ui{\mathbbm{i}}
\global\long\def\ud{\mathrm{d}}

\title{Supplemental Material for ``Impurity-induced thermal crossover in fractional Chern insulators''}

\author{Ke Huang}
\affiliation{Department of Physics, City University of Hong Kong, Kowloon, Hong Kong SAR, China}

\author{Sankar Das Sarma}
\affiliation{Condensed Matter Theory Center and Joint Quantum Institute, University of Maryland, College Park, Maryland 20742, USA}
\affiliation{Kavli Institute for Theoretical Physics, University of California, Santa Barbara, California 93106, USA}

\author{Xiao Li}
\email{xiao.li@cityu.edu.hk}
\affiliation{Department of Physics, City University of Hong Kong, Kowloon, Hong Kong SAR, China}

\date{\today}

\maketitle

\setcounter{equation}{0} 
\setcounter{figure}{0}
\setcounter{table}{0} 
\renewcommand{\theparagraph}{\bf}
\renewcommand{\thefigure}{S\arabic{figure}}
\renewcommand{\theequation}{S\arabic{equation}}
\renewcommand\labelenumi{(\theenumi)}

\maketitle

\section{The Projected Hamiltonian}

In this section, we present the projected Hamiltonian in the main text.
After projecting the interaction onto the highest valence band, we obtain
\begin{align}
	P_{\text{flat}}H_{\text{int}}P_{\text{flat}}
	&=\frac1{2A}\sum_{\vb*k,\vb*k'\vb*q\in\text{MBZ}}U(\vb*k,\vb*k',\vb*q)(f^\dag_{\vb*k+\vb*q}f_{\vb*k}-\delta_{\vb*q,0}/2)(f^\dag_{\vb*k'-\vb*q}f_{\vb*k'}-\delta_{\vb*q,0}/2),
\end{align}
where $f_{\vb*k+\vb*q}$ is the annihilation operator of a particle with momentum $\vb*k$ in the highest valence band. The matrix element $U(\vb*k,\vb*k',\vb*q)$ is calculated by
\begin{align}
	U(\vb*k,\vb*k',\vb*q)=\sum_{\vb*G\in\Lambda}V(\abs{\vb*q+\vb*G})\mel{\psi_{\vb*k+\vb*q}}{e^{i(\vb*q+\vb*G)\vdot\hat{\vb*r}}}{\psi_{\vb*k}} 
	\mel{\psi_{\vb*k'-\vb*q}}{e^{-i(\vb*q+\vb*G)\vdot\hat{\vb*r}}}{\psi_{\vb*k'}},
\end{align}
where $\Lambda=\{m\vb*g_1+n\vb*g_2: (m,n)\in\mathbb{Z}^2\}$ is the set of all reciprocal lattice vectors, and $\ket{\psi_{\vb*k}}$ is the eigenstate of the highest valence band.

The projected impurity potential is given by 
\begin{align}
	P_{\text{flat}}H_{\text{int}}P_{\text{flat}}=V_{\text{imp}}\sum_{i=1}^{N_{\text{imp}}}\sum_{\vb*k,\vb*k'}\mel{\psi_{\vb*k}}{\delta(\hat{\vb*r}-\vb*r_i)}{\psi_{\vb*k'}}f_{\vb*k}^\dag f_{\vb*k'}/\expval{n(\vb*r_i)}_{\text{full}}, 
\end{align}
where the density of the completely filled band is calculated by $\expval{n(\vb*r_i)}_{\text{full}}=\sum_{\vb* k}\mel{\psi_{\vb*k}}{\delta(\hat{\vb*r}-\vb*r_i)}{\psi_{\vb*k}}$.

\section{Calculation of the particle/hole entanglement spectrum}

In this section, we discuss how to calculate the particle-entanglement spectrum (PES) and hole-entanglement spectrum through $n$-point correlation functions. 
For a many-body state containing $n_e$ particles, we can divide them into two subsystems: one with $n_a$ particles and the other with $n_b=n_e-n_a$ particles. 
Supposing that the antisymmetrized many-body wave function is $\Psi(x_1,x_2,\cdots,x_{n_e})$, the reduced density matrix by tracing out $n_b$ particles is defined as
\begin{align}
	\rho(x_1,x_2,\cdots,x_{n_a},y_1,y_2,\cdots,y_{n_a})
	&=\int\dd{r_1}\dd{r_2}\cdots\dd{r_{n_b}}\Psi(x_1,\cdots,x_{n_a},r_1,\cdots,r_{n_b})^*\nonumber\\
	&\quad\times\Psi(y_1,\cdots,y_{n_a},r_1,\cdots,r_{n_b}).
\end{align}
The PES is then defined as $\xi=-\ln \lambda$, where $\lambda$ is the eigenvalue of the reduced density matrix.
To diagonalize the density matrix, we express the density matrix in terms of $n_a$-point correlation function,
\begin{align}
	\rho(x_1,\cdots,x_{n_a},y_1,\cdots,y_{n_a})
	&=\frac{n_b!}{n_e!}\expval{\psi^\dag(x_{n_a})\cdots\psi^\dag(x_{1})\psi(y_{1})\cdots\psi(y_{n_a})},
\end{align}
where $\psi(x)$ is the field operator. Note that this definition also applies to mixed states. For a given basis $u_\alpha(x)$, the field operator can be decomposed into
\begin{align}
	\psi(x)=\sum_\alpha u_\alpha(x)c_a,
\end{align}
and the $n_a$-point correlation function becomes
\begin{align}
	\rho(x_1,\cdots,x_{n_a},y_1,&\cdots,y_{n_a}) 
	=\binom{n_e}{n_a}^{-1}\sum_{\alpha_1<\cdots<\alpha_{n_a}}\sum_{\beta_1<\cdots<\beta_{n_a}}
	\expval{c_{\alpha_{n_a}}^\dag\cdots c_{\alpha_{1}}^\dag  c_{\beta_{1}}\cdots c_{\beta_{n_a}}}\nonumber\\
	&\times \frac{u_{[\alpha_1}(x_1)^*\cdots u_{\alpha_{n_a}]}(x_{n_a})^*}{\sqrt{n_a!}} \, \frac{u_{[\beta_1}(y_1)\cdots u_{\beta_{n_a}]}(y_{n_a})}{\sqrt{n_a!}},
\end{align}
where the square parenthesis denotes the antisymmetrization of the indices within it. 
The wave functions in the second line are normalized Slater determinants of the single-particle basis states, so they form a complete basis for $n_a$-body states.
Hence, the spectrum of the density matrix is identical to the spectrum of 
\begin{align}
	[\rho]_{\alpha,\beta} =\binom{n_e}{n_a}^{-1} \expval{c_{\alpha_{n_a}}^\dag\cdots c_{\alpha_{1}}^\dag  c_{\beta_{1}}\cdots c_{\beta_{n_a}}},
\end{align}
where $\alpha,\beta$ denote the collection of $\alpha_{1}<\cdots<\alpha_{n_a}$ or $\beta_{1}<\cdots<\beta_{n_a}$, respectively. Similarly, to partition $n_h$ holes and obtain the hole entanglement spectrum, we need to diagonalize
\begin{align}
	[\rho]_{\alpha,\beta} =\binom{n_h}{n_a}^{-1} \expval{c_{\alpha_{n_a}}\cdots c_{\alpha_{1}}  c_{\beta_{1}}^\dag\cdots c_{\beta_{n_a}}^\dag}.
\end{align}

\begin{figure*}[t]
	\center
	\includegraphics[width=\textwidth]{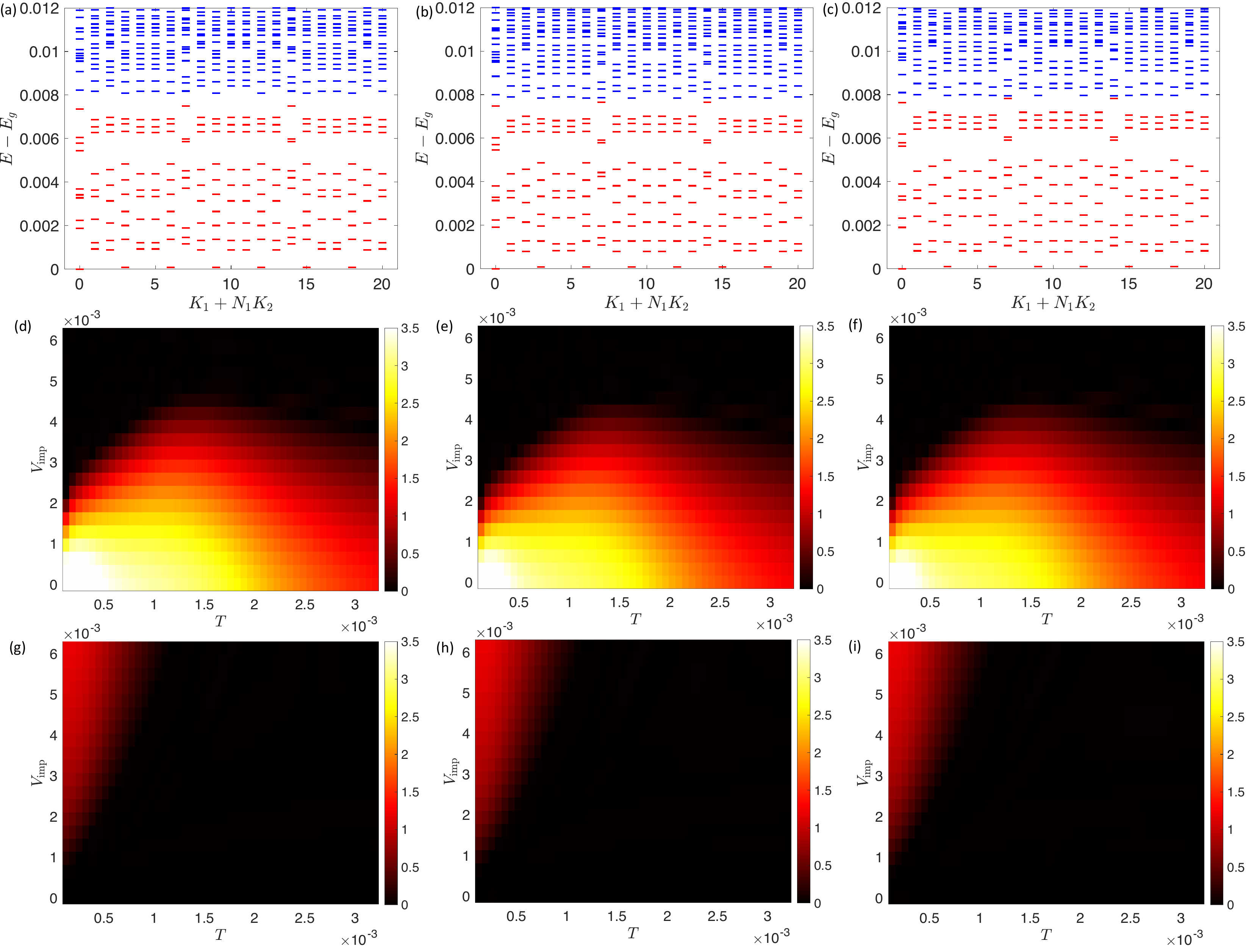}
	\caption{\label{Fig:w}(a-c) Enenergy spectrum for $w=0.9w_{\text{chiral}}, w_{\text{chiral}},1.1w_{\text{chiral}}$. (d-f) FCI entanglement gap for $w=0.9w_{\text{chiral}}, w_{\text{chiral}},1.1w_{\text{chiral}}$. (g-i) WC entanglement gap for $w=0.9w_{\text{chiral}}, w_{\text{chiral}},1.1w_{\text{chiral}}$. The other parameters are the same as those in the main text. The hole-entanglement spectrum is obtained by retaining three holes. For a pinned WC, there are $\binom{6}{3}=20$ states below the entanglement gap, and for an FCI, there are 637 states below the entanglement gap according to the (1,3)-permissible rule.
	}
\end{figure*}

\section{The effect of $w$ in the CTBG model on the crossover}

In this section, we study how a variation of the parameter $w$ in the CTBG model introduced in the main text affects the crossover between the pinned hole-Wigner crystal (WC) and the fractional Chern insulator (FCI). 
In Fig.~\ref{Fig:w}(a-c), we calculate the momentum-resolved many-body spectrum for $w=0.9w_{\text{chiral}}, w_{\text{chiral}},1.1w_{\text{chiral}}$, respectively. 
All three systems manifest a low-energy manifold of 196 states, corresponding to the quasiparticle excitations of the FCI state, but the energy gap above it is slightly greater for $w=0.9w_{\text{chiral}}$. 
Nonetheless, the entanglement gap for the three cases is nearly identical, as shown in Fig.~\ref{Fig:w}(d-i), suggesting the existence of such a crossover for all three $w$. 
Hence, we conclude that $w$ does not play an essential role as long as $w\approx w_{\text{chiral}}$.

\section{The effect of different $n_h$ and $N_{\text{imp}}$ on the crossover}

\begin{figure*}[t]
	\center
	\includegraphics[width=\textwidth]{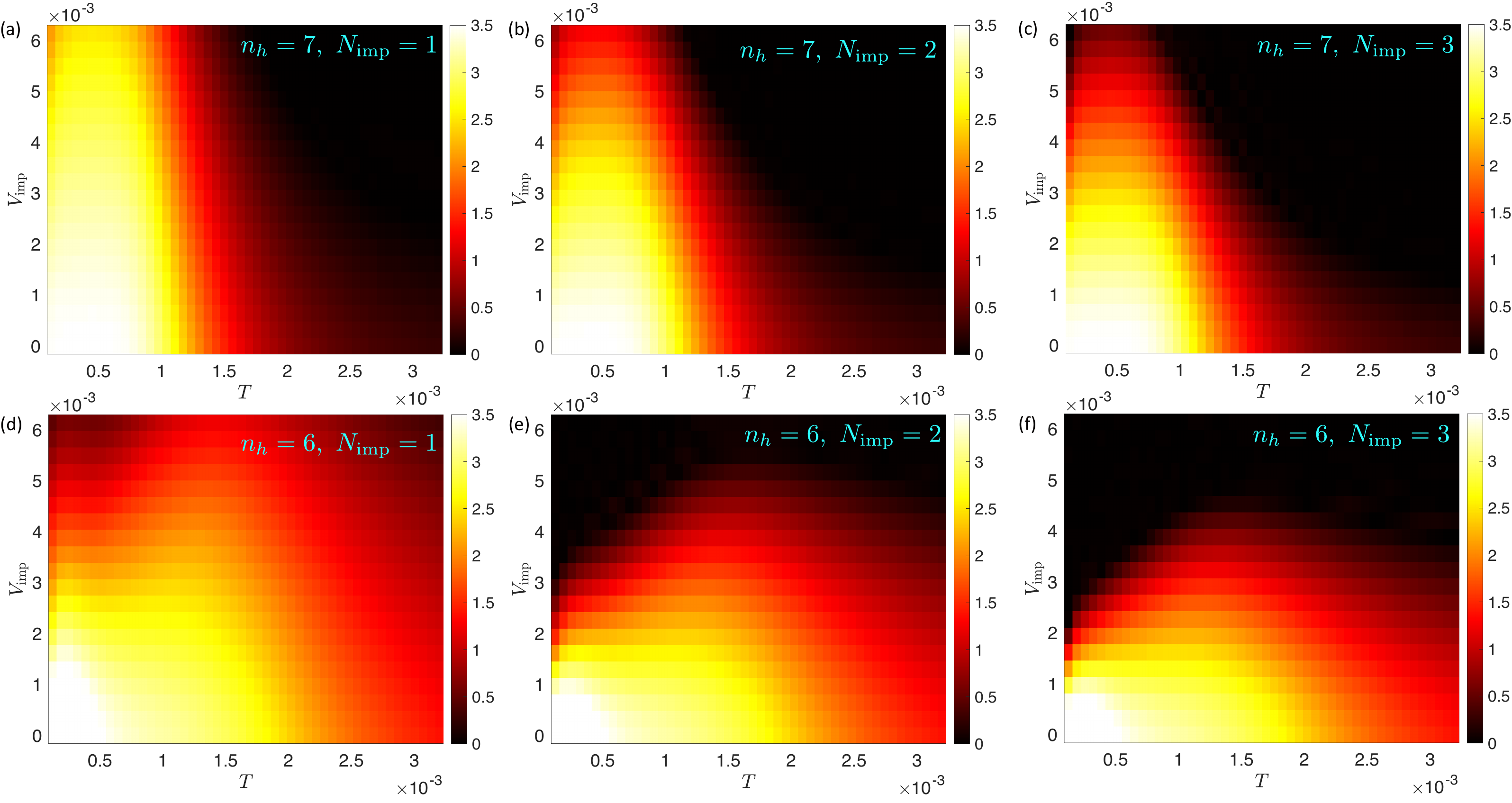}
	\caption{\label{Fig:EE}FCI entanglement gap in a system with 21 unit cells. Here, we retain 3 holes and trace out the rest of the holes in the calculation of the HES. For an FCI, there are 637 states below the entanglement gap according to the (1,3)-permissible rule.
	}
\end{figure*}

In Fig.~\ref{Fig:EE}(a-f), we present the FCI entanglement gap for six different cases. 
The results show several important differences between the cases with an exact 1/3 hole filling [Fig.~\ref{Fig:EE}(a-c), denoted as Group~A] and those with 1/3 filling minus one hole [Fig.~\ref{Fig:EE}(d-f), denoted as Group~B]. 
First, we observe that for Group~A, when the ground state does not have an FCI entanglement gap, raising the temperature will not induce an FCI entanglement gap either. 
This is because the entropy of the FCI state at exactly 1/3 filling is very low, and thus we do not expect an entropy-induced crossover at finite temperatures. 
Second, we observe that for a given impurity strength, the FCI entanglement gap for Group~B generally persists to a higher temperature than that for Group~A. 
In other words, $T_\text{FCI}$ (as defined in the main text) is larger for Group~B than for Group~A. 
This difference in $T_\text{FCI}$ can be attributed to entropy considerations as well, although the mechanism differs slightly from the one discussed in the main text, as we explain now.
The phase transition across $T_\text{FCI}$ can be understood by the following free-energy argument: $\Delta U - T \Delta S <0$, where 
$\Delta U$ is the energy penalty associated with exciting electrons from the low-energy FCI manifold to higher-energy states, and $\Delta S$ is the corresponding entropy gain.  
Therefore, a larger $\Delta S$ will induce a lower $T_\text{FCI}$ if $\Delta U$ is similar. 
For Group~A, the entropy of the FCI state is very low, and thus the entropy gain from exciting electrons to higher-energy states is large. 
Meanwhile, for Group~B, the entropy of the FCI state is much larger due to the abundance of quasiparticle excitations, and thus the entropy gain from exciting electrons to higher-energy states is smaller. 
These two results corroborate our entropy argument in the main text. 
Consequently, there is no crossover at the exact 1/3 filling in this toy model. 
However, we emphasize that the scenario may be different for the fractional quantum Hall states, whose neutral excitation may contribute a large entropy at the exact fractional fillings~\cite{Platzman1993}.
Third, we notice in Fig.~\ref{Fig:EE}(d) that one impurity is not sufficient to destroy the FCI state in the presence of quasiparticles, suggesting the robustness of FCI states in this toy model against small filling deviations. Finally, we only observe a well-developed WC entanglement gap for $n_h=6$ and $n_{\text{imp}}=3$ case, while the WC entanglement gaps are insignificant for all the other cases.


\begin{figure*}[t]
	\center
	\includegraphics[width=\textwidth]{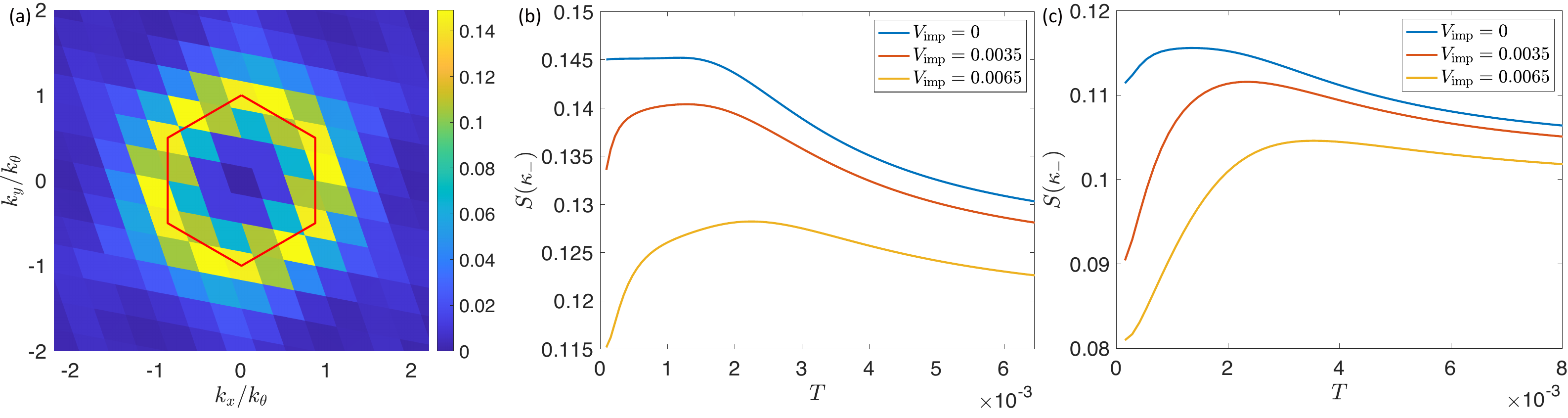}
	\caption{\label{Fig:SF}(a) CSF of the FCI ground state at the exact 1/3 hole filling without impurities. (b) CSF at $\kappa_-$ and finite temperatures in the system with finite impurity strength and at the exact 1/3 hole filling. (c) CSF at $\kappa_-$ and finite temperatures in the system with finite impurity strength and at the exact 1/3 hole filling minus one hole. 
	}
\end{figure*}

\section{The evaluation of the structure factor}

In this section, we study the connected structure factor (CSF) as a function of temperature. The CSF is defined as
\begin{align}
	S(\vb*q)=\frac1N\qty(\expval{\rho_{\vb*q}\rho_{-\vb*q}}-\expval{\rho_{\vb*q}}\expval{\rho_{-\vb*q}}),
\end{align}
where $N$ is the system size (number of unit cells). 
In Fig.~\ref{Fig:SF}(a), we calculate the CSF of the FCI ground state at the exact $1/3$ hole filling. 
The CFI of the FCI state features a relatively uniform ring around the boundary of the moir\'e BZ, which has also been observed in Ref.~\cite{Dong2023c}. 
Hence, we evaluate the CSF at $\kappa_-$ at different temperatures and impurity strengths in Fig.~\ref{Fig:SF}(b). 
Without impurities, the CFS manifests a plateau at low temperatures and decreases for $T>0.002$. 
For $V_{\text{imp}}=0.0035$, except for a small dip at zero temperature, the CFS behaves similarly to the clean limit, which agrees with the entanglement gap in Fig.~\ref{Fig:EE}(c). 
For $V_{\text{imp}}=0.0065$, where the FCI entanglement gap becomes very small, the CFS is lower at zero temperature and does not possess a plateau at low temperatures. 
We further calculate the CSF at $\kappa_-$ at the exact $1/3$ hole filling minus one hole. 
Even with three quasiparticles, the CFS in the clean limit also shows a plateau at zero temperature and decreases for $T>0.003$, resembling the result at the exact $1/3$ hole filling. 
However, the CFS becomes much lower at zero temperature for $V_{\text{imp}}=0.0035$, because the ground state becomes a hole-WC. 
Further increasing the temperature, the CFS peaks within $0.001<T<0.003$, consistent with the temperature window for FCI entanglement gap in Fig.~\ref{Fig:EE}(f), before it eventually decreases at higher temperatures. 
For $V_{\text{imp}}=0.0065$, the CFS becomes even lower at zero temperature with a less prominent peak.

\bibliographystyle{apsrev4-2}
\bibliography{FCI_dis}